\documentclass[aps,prc,twocolumn,showpacs,preprintnumbers,
                          nofootinbib,float,longbibliography]{revtex4-1}
\usepackage{graphicx, fancybox}
\usepackage[caption=false]{subfig}
\usepackage{amsmath,amssymb}
\usepackage[usenames, dvipsnames]{color}
\usepackage{soul}
\usepackage{url}

\usepackage{adjustbox}
\usepackage{slashed}
\usepackage{bm}
\usepackage{relsize}
\usepackage{color}

\usepackage[normalem]{ulem}

\newcommand{\RE}{\mathrm{Re}\,}
\newcommand{\IM}{\mathrm{Im}\,}
\newcommand{\ret}{\mathrm{ret}}
\newcommand{\adv}{\mathrm{adv}}

\newcommand{\beq}{\begin{equation}}
\newcommand{\eeq}{\end{equation}}

\usepackage{filecontents}

\begin{document}

\title{Photon emission from quark-gluon plasma out of equilibrium}

\author{Sigtryggur Hauksson }
\thanks{Corresponding author: sigtryggur.hauksson@mail.mcgill.ca}
\author{Sangyong Jeon}
 \author{Charles Gale}
 \affiliation{Department of Physics, McGill University, 3600 University
 Street, Montreal,
 QC, H3A 2T8, Canada}

\begin{abstract}
The photon emission from a non-equilibrium quark-gluon plasma (QGP) is analyzed. We derive an integral equation that describes photon production through quark-antiquark annihilation and quark bremsstrahlung. It includes coherence between different scattering sites, also known as the Landau-Pomeranchuk-Migdal effect. These leading-order processes are studied for the first time  together in an out-of-equilibrium field theoretical  treatment that enables the inclusion of viscous corrections to the calculation of electromagnetic emission rates. In the special case of an isotropic, viscous, plasma the integral equation only depends on three constants which capture the non-equilibrium nature of the medium. 
\end{abstract}

\maketitle
\date{\today}

\section{Introduction}

Relativistic collisions of large nuclei allow the study of QCD (Quantum Chromodynamics: the theory of the nuclear strong interaction) matter at high temperatures. Experiments performed at the Relativistic Heavy Ion Collider (RHIC) and the Large Hadron Collider (LHC) have indeed shown that such collisions create droplets of quark-gluon plasma (QGP) \cite{Jacak:2012dx}.  A significant breakthrough in the relativistic heavy-ion program has been the realization that this fluid and its evolution can be characterized by relativistic hydrodynamics, which describes long-wavelength excitations \cite{[{See, for example, }][{, and references therein.}]Gale:2013da}.  Therefore, those experiments have the potential to give access to the transport coefficients of QGP, such as shear and bulk viscosity. These coefficients are fundamental properties of QCD. 

There has been extensive work on extracting the viscosity of QGP from soft hadronic observables \cite{[{See, for example, }][{, and references therein.}]Heinz:2013th}. In addition, electromagnetic observables, i.e. photons and dileptons, can play an important role in that endeavour \cite{Vujanovic:2013jpa,Paquet:2015lta}. They are emitted throughout the evolution of the QGP and escape from the medium unaffected by final state interaction. This work will focus on the production of real photons. 

In order to evaluate the effects of viscosity on photons one must study their out-of-equilibrium emission. 
At leading order in the strong coupling constant there are two channels for photon production in QGP. Firstly, there are two-to-two scattering channels with a photon in the final state. They were first calculated in \cite{Baier1991,Kapusta1991} for thermal equilibrium. Since then there have been a number of works on these processes in an out-of-equilibrium QGP such as  for finite fugacities \cite{Strickland1994,Baier1997,Gelis2004},  in an anisotropic QGP \cite{Schenke2006b},  for shear viscous corrections  \cite{Dion:2011pp,Shen2014} and  for bulk viscous corrections \cite{Hauksson2016}. Note that consistent calculations of the electromagnetic emissivity should include non-equilibrium corrections to the thermal mass of the soft mediators.

\begin{figure}
\begin{center}
	\includegraphics[width=0.95\columnwidth]{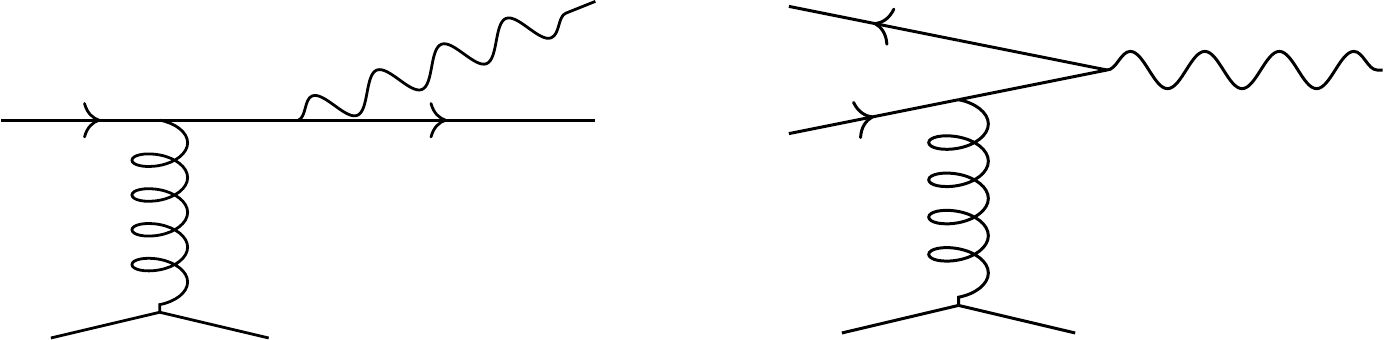}
\end{center}
	\caption{Photon production through quark bremsstrahlung and the annihilation of a quark and an antiquark. For the first process, the diagram with the photon emitted left of the vertex is included but not shown, as is that with the gluon connecting to the antiquark in the second process.}
	\label{fig:bremsstrahlung}
\end{figure}

Secondly, there are inelastic channels with bremsstrahlung off a quark and the pair annihilation of a quark and antiquark, see Fig. \ref{fig:bremsstrahlung}. These processes contribute as much to photon emission as two-to-two scattering \cite{Aurenche2000}. 
 The photon is emitted almost collinearily to the quark, which entails a large decoherence time. This means that the quarks can exchange arbitrarily many soft gluons with the medium during the formation of the photon. This leading-order complication is known as the Landau-Pomeranchuk-Migdal (LPM) effect \cite{Landau:1953um,Landau:1953gr,Migdal:1955nv,Migdal:1956tc}. It means that in addition to the diagrams in Fig. \ref{fig:bremsstrahlung} one must sum up diagrams with an arbitrary number of gluon exchanges such as in Fig. \ref{fig:LPM4}. The LPM effect was first treated consistently in \cite{Arnold2001,Arnold2001b} for a medium in thermal equilibrium where it was shown to reduce photon production because of coherence between different emission sites.

This work treats the photon production through inelastic channels in a non-equilibrium QGP for the first time, using a field theoretical derivation which includes the LPM effect without relying on the Kubo-Martin-Schwinger (KMS) \cite{Kubo:1957mj,*Martin:1959jp} relation which describes detailed balance in thermal equilibrium.  
The field-theoretical techniques used here differ from  those employed in Ref. \cite{AMY_eff_kin}, where kinetic theory was used to analyze inelastic scattering of quarks and gluons.

The paper is organized as follows. In Sec. \ref{Resummed} we review the real-time formalism and derive expressions for resummed propagators. In Sec. \ref{Soft_gluons} we discuss the resummed \(rr\) propagator for soft gluons and show that the LPM effect is leading order in non-equilibrium systems. Sec. \ref{Hard_quarks} discusses the resummed occupation number of hard quarks. In Sec. \ref{Sum} we sum up the different diagrams contributing to the LPM effect and derive an integral equation describing the inelastic channels. Finally, we conclude in Sec. \ref{Conclusion} and indicate future directions.    We will use the \((+,-,-,-)\) metric. If \(P^{\mu}\) is a four-vector we write \(P^{\mu} = (p^0,\mathbf{p})\) and define \(p = |\mathbf{p}|\) and \(\mathbf{\hat{p}} = \mathbf{p}/p\).

\section{Resummed propagators in the real-time formalism} \label{Resummed}

\begin{figure}               
\centering
	\includegraphics[width=0.6\columnwidth]{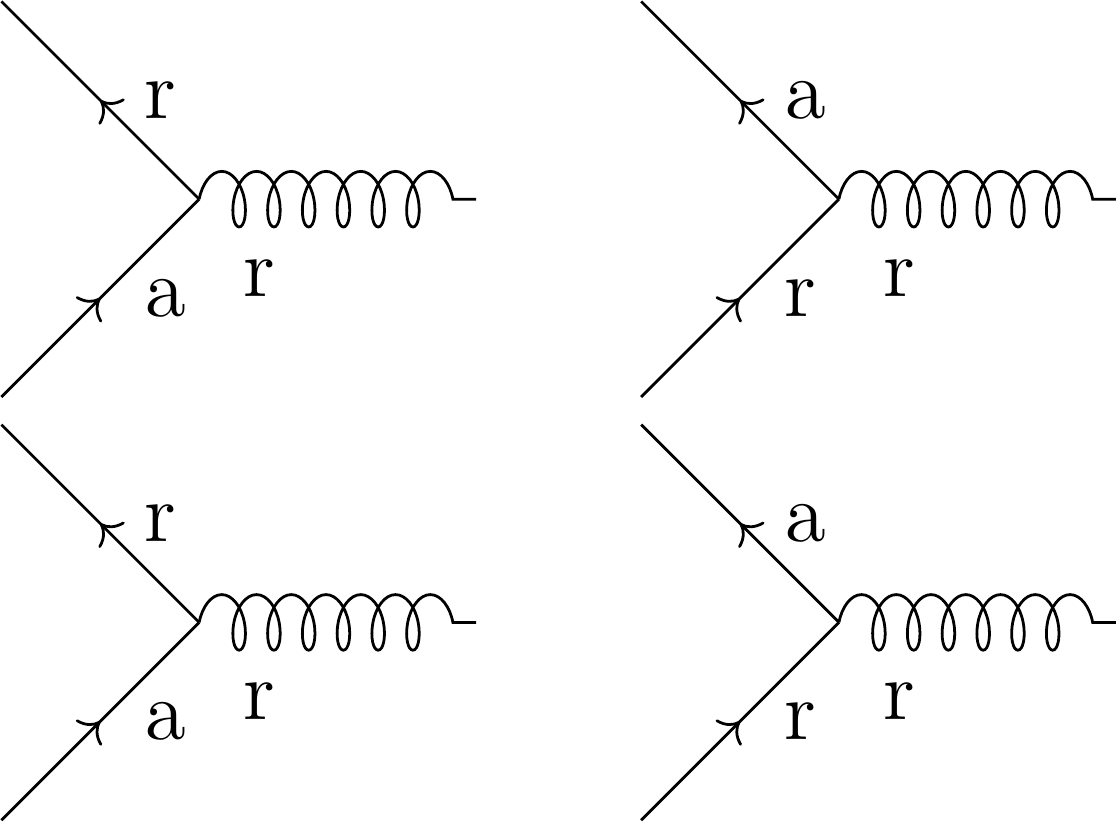}
	\caption{Quark-gluon vertices in the $r/a$ basis in the real time formalism}
	\label{fig:ravertices}
\end{figure}

Out-of-equilibrium quantum field theorie is best described in the real-time formalism where a closed time contour leads to the doubling of degrees of freedom \cite{Bellac2011, Chou1984}. In this paper we mostly work in the \(r/a\) basis \cite{Keldysh1964} which is defined by 
\beq 
\phi_r = \frac{1}{2}\left( \phi_1 + \phi_2\right), \quad \phi_a = \phi_1 - \phi_2.
\eeq  
The propagators are 
\beq \label{eq:prop_def}
D_{cd}(x,y) = \langle \phi_c(x)\phi^{\dagger}_d(y)\rangle = \mathrm{Tr}\left[ \rho_0 \;\mathcal{T}_{\mathcal{C}} \phi_c(x)\phi^{\dagger}_d(y)\right]
\eeq 
where \(c\) and \(d\) are either \(r\) or \(a\). The initial density matrix \(\rho_0\) determines the out-of-equilibrium evolution of the system. In the \(r/a\) basis, vertices have an odd number of \(a\) indices, see Fig. \ref{fig:ravertices}. This basis has numerous advantages: The \(aa\) propagator vanishes identically and the bare \(ra\) and \(ar\) propagators only include vacuum contributions. Furthermore it allows for easier power counting.  For concreteness we consider complex scalar fields in this section but our arguments can easily be generalized. 

In this paper we study translationally invariant systems. In other words, our calculation is for a homogeneous and static brick of out-of-equilibrium QGP. This approximation is justified when the mean free path of quasiparticles is much smaller scale than the macroscopic scale at which hydrodynamical quantities change appreciably. In a translationally invariant system the propagators are 
\begin{align}
\begin{split} 
D_{rr}(x) &= \frac{1}{2}\langle\{\phi(x), \phi^{\dagger}(0) \} \rangle \\
D_{ra}(x) &=  \theta(x^0) \left\langle \left[\phi(x), \phi^{\dagger}(0)\right] \right\rangle\\
D_{ar}(x) &=  - \theta(-x^0) \left\langle \left[\phi(x), \phi^{\dagger}(0)\right] \right\rangle \\  
D_{aa}(x) &= 0.
\end{split}
\end{align}
We see that \(D_{ra} = D_{\ret}\) and \(D_{ar} = D_{\adv}\). Going to momentum space with momentum \(P\)
\begin{align} 
\begin{split}
D_{\mathrm{ret}}(P)^{*} &= \int d^4 x \; e^{-iP \cdot x} \theta(x^0) \left\langle \left[\phi(0), \phi^{\dagger}(x)\right] \right\rangle \\
&= \int d^4 x \; e^{iP \cdot x} \theta(-x^0) \left\langle \left[\phi(0), \phi^{\dagger}(-x)\right] \right\rangle \\
&= \int d^4 x \; e^{iP \cdot x} \theta(-x^0) \left\langle \left[\phi(x), \phi^{\dagger}(0)\right] \right\rangle \\
&= - D_{\mathrm{adv}}(P)
\end{split}
\end{align} 
where  we did a change of variables \(x \rightarrow -x\) in the second line and then used translational invariance. This shows that the resummed propagators only have two independent components. The bare retarded propagator can easily be evaluated because the commutator of free bosonic fields is just a number so summing over all states is trivial. It is the same as in vacuum,
\beq \label{eq:bare_ret_adv}
D^0_{\ret}(P) = \frac{i}{P^2+i\epsilon p^0}
\eeq

We have yet to find the \(rr\) propagator. In thermal equilibrium the Kubo-Martin-Schwinger (KMS) relation stipulates that 
\beq \label{eq:rr_equilibrium}
D_{rr} (P) =  \left( \frac{1}{2} + f_B(p^0)\right) \left[D_{\mathrm{ret}}(P) - D_{\mathrm{adv}}(P) \right]
\eeq
so there is only one independent propagator. Here \(f_B(p^0)\) is the Bose-Einstein distribution \cite{Bellac2011}. This expression is valid at every order in perturbation theory and thus offers great simplification. 
 Using Eq. \eqref{eq:bare_ret_adv} the bare \(rr\) propagator is 
\beq 
D^0_{rr} (P) =  \left( \frac{1}{2} + f_B(p)\right) 2\pi \delta(P^2).
\eeq

In non-equilibrium systems one cannot obtain a general expression for the resummed \(rr\) propagator. In analogy with the equilibrium case the bare propagator is  
\beq \label{eq:bare_rr_non_eq}
D^0_{rr}(P) = \left( \frac{1}{2} + \theta(p^0)f(\mathbf{p}) + \theta(-p^0)f(-\mathbf{p})\right) 2\pi \delta(P^2)
\eeq 
where the ansatz, \(f(\mathbf{p})\), is a general momentum distribution characterizing the system. For mirror symmetric momentum distributions, \(f(\mathbf{p}) = f(-\mathbf{p})\), the bracket reduces to \(1/2+f(\mathbf{p})\), see \cite{Mrowczynski1992} for a discussion. Eq. \eqref{eq:bare_rr_non_eq} can be justified from first principles as in
\cite{Calzetta1986,Berges2004}.
Assuming an initial density matrix \(\rho_0\) one Legendre-transforms the path integral from external sources to connected n-point functions. This gives rise to an infinite tower of equations corresponding to the BBGKY hierarchy in kinetic theory. Truncating the tower at second order and assuming that propagators vary slowly in space one gets Eq. \eqref{eq:bare_rr_non_eq} at lowest order in the coupling. The function \(f\) can be shown to be real and positive and the equations of motion reduce to a Boltzmann equation for \(f\). Therefore, at leading order in the coupling constant  it should be interpreted as a momentum distribution of particles. The delta function shows that the quasiparticles are on shell to lowest order. In this approach \(f\) is left unspecified and can be chosen to match a hydrodynamical evolution of the QGP.  

Demanding that propagators vary slowly in time sets constraints on how far from equilibrium one can go. In anisotropic and translationally invariant systems the retarded gluon propagator acquires a pole with \(\mathrm{Im}\: \omega >0\) \cite{Romatschke2003}. This pole signals the exponential growth of the occupation density of soft gluons \cite{Mrowczynski2016} which can invalidate our assumption of translational invariance. 
Specifically, the pole introduces divergences in momentum integrals over \(G_{rr}\) for soft gluons. Throughout our analysis we will assume that the anisotropy is small enough so that this divergence does not appear at leading order in the coupling. 
As an example consider a  momentum distribution of the form \cite{Romatschke2003}
\beq 
\label{eq:RS_mom_distr}
f(\mathbf{p}) = f_{\mathrm{eq}}\left( \sqrt{p^2 + \xi (\mathbf{n}\cdot\mathbf{p})^2} \right)
\eeq
where \(f_{\mathrm{eq}}\) is an equilibrium distribution and \(\mathbf{n}\) is a unit vector specifying the direction of the anisotropy \(\xi\) (in general $-1 \leq \xi < \infty$). 
We will show below that we must demand \( | \xi | \lesssim g^2\) if the divergence is to be subleading. On the contrary, isotropic systems can be much further away from equilibrium without our analysis breaking down. In summary, we study systems with low anisotropy that are close enough to local thermal equilibrium so that the equilibrium power counting scheme is unaltered. This guarantees that the hard thermal loop (HTL) scheme remains valid.

In this paper we will need resummed propagators in out-of-equilibrium systems.  For the convenience of the reader we reproduce some known results for scalar field theory in the real-time formalism, see \cite{Greiner1998}. 
The Dyson equation is  
\begin{widetext}
\begin{equation} \label{eq:Dyson}
\begin{bmatrix}
D_{rr} & D_{ra} \\
D_{ar} & D_{aa}
\end{bmatrix}
=
\begin{bmatrix}
D_{rr}^0 & D_{ra}^0 \\
D_{ar}^0 & D_{aa}^0
\end{bmatrix}
+
\begin{bmatrix}
D_{rr}^0 & D_{ra}^0 \\
D_{ar}^0 & D_{aa}^0
\end{bmatrix}
(-i)
\begin{bmatrix}
\Pi_{rr} & \Pi_{ra} \\
\Pi_{ar} & \Pi_{aa}
\end{bmatrix}
\begin{bmatrix}
D_{rr} & D_{ra} \\
D_{ar} & D_{aa}
\end{bmatrix}.
\end{equation}
\end{widetext}
where, say, \(\Pi_{aa}\) is the self-energy sourced by two \(a\) fields. Using \(D_{aa} = D^0_{aa} = 0\) one finds that \(\Pi_{rr} = 0\). Defining \(\Pi_{ar} = \Pi_{\mathrm{ret}}\) one obtains
\beq \label{eq:Dyson_ret}
D_{\mathrm{ret}} = D^0_{\mathrm{ret}} + D^0_{\mathrm{ret}} \left(-i\Pi_{\mathrm{ret}}\right) D_{\mathrm{ret}} 
\eeq
which gives
\beq \label{eq:resummed_ret}
D_{\mathrm{ret}} = \frac{i}{P^2 -  \Pi_{\mathrm{ret}}}.
\eeq
This equation gives the dispersion relation for the quasi-particles, i.e. their thermal mass and decay width. Similarly
\beq \label{eq:resummed_adv}
D_{\mathrm{adv}} = \frac{i}{P^2 -  \Pi_{\mathrm{adv}}}
\eeq 
where \(\Pi_{\mathrm{adv}} = \Pi_{ra} = \Pi_{\mathrm{ret}}^{*}\) in translationally invariant systems.

The Dyson equation for \(D_{rr}\) is more complicated. Defining \(D_< = D_{12}\) and \(D_> = D_{21}\) we can write 
\beq \label{eq:D_rr_and_D_<} 
D_{rr} = \frac{1}{2} \left(D_{>} + D_{<} \right) = D_{<} + \frac{1}{2} \left(D_{\mathrm{ret}} - D_{\mathrm{adv}} \right)
\eeq
since \(D_> - D_< = D_{\ret} - D_{\adv}\). We will analyze \(D_{<}\) to obtain equations with a clear physical interpretation.
Using the \(rr\) component of Eq. \eqref{eq:Dyson}, Eq. \eqref{eq:Dyson_ret} and the corresponding equation for \(D_{\adv}\) one gets
\begin{widetext}
\begin{equation}
D_{<} = D_{<}^0 +  D_{\mathrm{ret}}^0 \left(-i\Pi_{\mathrm{ret}}\right) D_{<} + D_{<}^0 \left(-i\Pi_{\mathrm{adv}}\right) D_{\mathrm{adv}} + D_{\mathrm{ret}}^0 \left(-i\Pi_{<}\right) D_{\mathrm{adv}}
\end{equation}
\end{widetext}
where
\begin{equation}
\Pi_{<} = \Pi_{aa} - \frac{1}{2}\Pi_{\mathrm{ret}} +\frac{1}{2}\Pi_{\mathrm{adv}} 
\end{equation} 
In the original \(12\) basis \(\Pi_{<} = - \Pi_{12}\) in our convention. This component of the self-energy describes the creation rate of quasi-particles \cite{Bellac2011}. 
Solving for \(D_{<}\) using Eq. \eqref{eq:resummed_ret} one gets  
\begin{equation}
D_{<} = \frac{\left(-iD^0_{\mathrm{ret}}\right)^{-1} D^0_< \left(-iD^0_{\mathrm{adv}}\right)^{-1} \quad + \quad i\Pi_<}{\left[ \left(-iD^0_{\mathrm{ret}}\right)^{-1} - \Pi_{\mathrm{ret}}\right] \left[\left(-iD^0_{\mathrm{adv}}\right)^{-1} - \Pi_{\mathrm{adv}}\right]}.
\end{equation}
For non-vanishing self-energy the first term is zero because
\beq 
D^0_< \left(D^0_{\mathrm{ret}}\right)^{-1} \propto P^2 \delta(P^2) = 0.
\eeq
This is true since our theory is defined in momentum space and assumes translational invariance. Thus
\begin{equation} \label{eq:D_<}
D_{<} = D_{\mathrm{ret}}\left(-i\Pi_<\right) D_{\mathrm{adv}}.
\end{equation}
and
\beq \label{eq:D_rr}
D_{rr} = \frac{1}{2} \left(D_{\ret} - D_{\adv} \right) + D_{\mathrm{ret}}\left(-i\Pi_<\right) D_{\mathrm{adv}}
\eeq
In deriving this equation we never had to invert the order of propagators. Thus it is equally valid for spinors and spin-1 bosons whose propagators are matrices in the spacetime indices.

For \(D_> = D_{21}\) one similarly gets that
\beq 
D_> = D_{\ret} \left( -i \Pi_>\right) D_{\adv}
\eeq
where \(\Pi_> = - \Pi_{21}\) describes the annihilation of quasi-particles. It's easy to see that
\beq 
\Pi_{\ret} - \Pi_{\adv} = \Pi_{>} - \Pi_<
\eeq
which reduces to 
\beq 
2i \IM \Pi_{\ret} = \Pi_{>} - \Pi_<
\eeq
in translationally invariant systems. This last equation says that the decay width of a quasi-particle is the difference of the annihilation and creation rate.  

For scalar particles we can go further and derive a more intuitive expression for \(D_{rr}\). In translationally invariant systems Eq. \eqref{eq:resummed_ret} and  \eqref{eq:resummed_adv} give that 
\begin{equation}
D_{\mathrm{ret}} - D_{\mathrm{adv}} = 2 \left( \IM \Pi_{\mathrm{ret}} \right) D_{\mathrm{ret}} D_{\mathrm{adv}}.
\end{equation}
Thus we see that
\begin{equation} \label{eq:scalar_rr_resummed}
D_{rr} = \left[ \frac{1}{2} + \frac{\Pi_<}{2i \IM \Pi_{\ret}} \right]\left(D_{\mathrm{ret}} - D_{\mathrm{adv}} \right).
\end{equation}
This equation has a striking resemblance with the \(rr\) propagator in equilibrium, Eq. \eqref{eq:rr_equilibrium}. Indeed \(\Pi_</2i \IM \Pi_{\ret}\) reduces to the Bose-Einstein distribution by using the KMS relation for self-energies. In non-equilibrium systems \(\Pi_</2i \IM \Pi_{\ret}\) is in general not the same as the bare momentum distribution \(f(\mathbf{p})\). It can be viewed as a resummed occupation density. We emphasize that we have only derived Eq. \eqref{eq:scalar_rr_resummed} for scalar particles since we needed to invert the order of propagators. In the next two sections we will derive a similar relation for soft gluons and hard quarks and evaluate the resummed occupation density explicitly.

\section{The \(rr\) propagator of soft gluons} \label{Soft_gluons}

The photon production rate is given by the \(12\) component of the photon polarization tensor
\beq \label{eq:photonrate}
k \frac{d R}{d^3k} = \frac{i}{2 (2\pi)^3} \left( \Pi^{\gamma}_{12} \right)^{\mu}_{\;\;\mu}\ ,
\eeq
where \(\mathbf{k}\) is the photon momentum and \(\Pi^{\gamma}_{12}\) is one component of the photon polarization tensor. This equation is valid in non-equilibrium systems as has been shown in \cite{Serreau2003}.

\begin{figure}
\centering
	\includegraphics[width=0.25\textwidth]{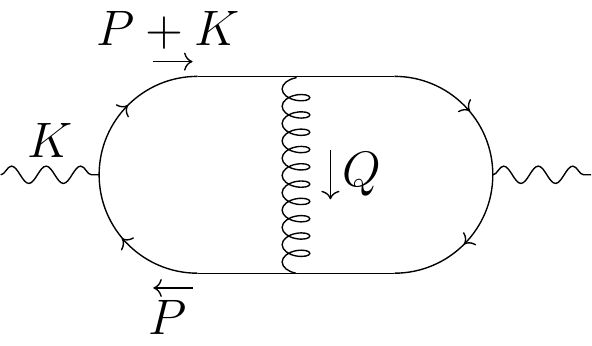}
	\caption{Definition of momenta in the argument for bremsstrahlung and pair annihilation contribution at leading order.}
    \label{fig:bremsstrahlung4}
\end{figure}

\begin{figure}
\centering
	\includegraphics[width=0.25\textwidth]{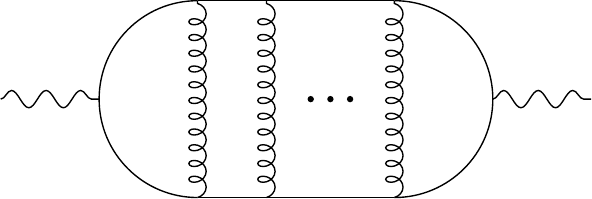}
	\caption{The diagrams for the LPM effect.}
    \label{fig:LPM4}
\end{figure}

The diagram corresponding to bremsstrahlung and quark-antiquark pair annihilation is in Fig. \ref{fig:bremsstrahlung4}. Due to the LPM effect the quarks can have arbitrarily many gluon exchanges, see Fig. \ref{fig:LPM4}. We will now briefly explain why these diagrams contribute at leading order for a medium in thermal equilibrium, see \cite{Aurenche1998,Arnold2001} for further details. The quarks are hard, \(P\sim T\), and nearly on shell, \(P^2 \sim g^2 T^2\), where \(T\) is the temperature and \(g \ll 1 \) is the strong coupling constant. The photon is emitted with an angle \(\theta \sim g\) relative to the quark momentum. Finally, the exchanged gluons are soft, \(Q \sim gT\), forcing us to use resummed propagators. 

We analyze the diagram in Fig. \ref{fig:bremsstrahlung4}. In thermal equilibrium the \(rr\)  propagator for soft gluons is 
\beq \label{eq:G_rr_eq}
G_{rr}(Q) = \left(\frac{1}{2} + f_B(q^0) \right) \left[ G_{\ret} - G_{\adv} \right] \sim \frac{1}{g^3T^2}
\eeq   
where \(f_B(q^0)\sim T/q^0 \sim 1/g\) and the retarded gluon propagator is \(G_{\ret} \sim 1/g^2 T^2\). Furthermore, each pair of quark propagators gives pinching poles of order \(1/g^2\). This can be seen more easily for bare scalars for which
\begin{align} \label{eq:pinching_poles}
\begin{split}
\int dp^0\;&D_{ar}(K+P) D_{ra}(P) \\
&= \int dp^0\; \frac{1}{\left[ (p^0 + i\epsilon)^2 - p^2\right] \left[(p^0 + k - i \epsilon)^2  - |\mathbf{p} + \mathbf{k}|^2\right]} \\
&\sim \frac{1}{T^2} \times \frac{1}{p+k-|\mathbf{p} + \mathbf{k}|} \sim \frac{1}{g^2T^3}
\end{split}
\end{align}
where we did a contour integration and used that \(\hat{\mathbf{p}} \cdot \hat{\mathbf{k}} = 1 - \mathcal{O}(g^2)\). In real calculations one must use resummed fermion propagators since their self-energy is \(\mathcal{O}(g^2)\). Finally each gluon vertex contributes a factor \(g\) and each photon vertex contributes a factor \(e\) as well as a factor \(g\) because of kinematics \cite{Arnold2001}. Including a \(g^3\) phase space suppression because \(\mathbf{q}\) is soft and a \(g^2\) suppression because \(\mathbf{p}\) is collinear with \(\mathbf{k}\) one sees that the diagram is of order \(g^2 e^2\). A similar analysis shows that Fig. \ref{fig:LPM4} is also leading order.

The above argument relied mostly on kinematics and is therefore equally valid in non-equilibrium systems. \footnote{In systems that are far away from thermal equilibrium there is no well defined temperature. Then the scale \(T\) should be replaced by the hard scale which contributes to the greatest number of particles, see \cite{AMY_eff_kin} for further details.}
 Nevertheless, it assumed thermal equilibrium in two crucial places. Firstly, the authors of \cite{Arnold2001} used a KMS condition for four-point functions to show that only \(S_{ra}\) and \(S_{ar}\) contribute to the pinching poles. We provide a more general argument in the next two sections. Secondly, Eq. \eqref{eq:G_rr_eq} for the \(rr\) propagator was derived using the KMS condition.

In general the retarded self-energy for soft gluons is \cite{Mrowczynski2000,AMY_eff_kin}
\begin{eqnarray}
\Pi_{\mathrm{ret}}^{\mu\nu} (Q) &= -2 g^2 \int \frac{d^3 p}{(2\pi)^3}\,  \frac{1}{2p} \left(\frac{\partial f_{\mathrm{tot}} (\mathbf{p})}{\partial P^{\omega}}  \right) \nonumber \\
&\times 
\left[- P^{\mu} g^{\omega\nu} +  \frac{Q^{\omega}P^{\mu}P^{\nu}}{P\cdot Q + i\epsilon}\right].  
\end{eqnarray}
where 
\beq
f_{\mathrm{tot}} = N_f f_q + N_f f_{\bar{q}} + 2N_c f_g
\eeq
with \(f_q\), \(f_{\bar{q}}\), \(f_g\) the distribution for quarks, antiquarks and gluons respectively.
We should interpret \(\partial f_{\mathrm{tot}} (\mathbf{p})/\partial p^{0} = 0\). 
The \(12\) component, \(\Pi_{<}(Q)\), has also been evaluated for space-like gluons \cite{AMY_eff_kin}.  It is 
\begin{align} \label{eq:Pi<}
\begin{split}
\Pi_{<}^{\mu\nu} &(Q) = -i g^2 \int \frac{d^3 p}{(2\pi)^3}\,  \frac{P^{\mu}P^{\nu}}{p}  2\pi \delta(P\cdot Q)\bigg|_{p^0 = p} \\
&\times \Big[ N_f f_q(\mathbf{p}) (1-f_q(\mathbf{p})) + N_f f_{\bar{q}}(\mathbf{p}) (1-f_{\bar{q}}(\mathbf{p})) \\
&\hspace{3.45cm} +  2 N_c f_g(\mathbf{p}) (1+f_g(\mathbf{p}))\Big].
\end{split}
\end{align} 
In both expressions we have used that \(P \gg Q\). Eq. \eqref{eq:Pi<} has an intuitive interpretation. The soft gluons are sourced by hard quarks and gluons with momentum P. The factor \(f_q(\mathbf{p}) (1-f_q(\mathbf{p}))\) is the density of the incoming and outgoing hard quark including Pauli blocking. Similarly  \(f_g(\mathbf{p}) (1+f_g(\mathbf{p}))\) describes the hard gluon.  These expressions are true as long as the hard thermal loop (HTL) scheme is valid. This puts some mild constraints on the momentum distribution  \cite{AMY_eff_kin}, such as that the density of soft gluons cannot be exceedingly high.

We can now see that 
\(G_{rr} \sim g^{-3}\) for soft gluons in non-equilibrium systems. 
Clearly \(\Pi_{\ret} \sim g^2 T^2\) so \(G_{\ret} \sim g^{-2}\) while \(\Pi_{<} \sim g T^2\) because of the delta function in Eq. \eqref{eq:Pi<}. Thus, using Eq. \eqref{eq:D_rr},
\beq 
G_{rr} \approx G_{\mathrm{ret}}\left(-i\Pi_<\right) G_{\mathrm{adv}} \sim \frac{1}{g^3 }
\eeq 
which ensures that the LPM effect matters at leading order.

\begin{figure}
\centering
	\includegraphics[width=0.18\textwidth]{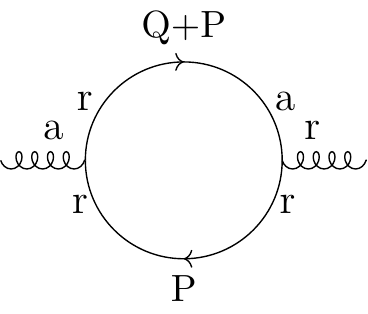}
	\caption{One of the diagrams contributing to $\Pi_{\ret} = \Pi_{ar}$}
    \label{fig:quarkloop_ra}
\end{figure}

At first sight it might be surprising that \(\Pi_{<}\) and \(\Pi_{\ret}\) differ by a power of \(g\) since they come from the same Feynman diagrams. The reason for the difference is the following: The vertices give factors \(P^{\mu}P^{\nu}\) which are \(\mathcal{O}(1)\) and factors \(P^{\mu}Q^{\nu} + Q^{\mu}P^{\nu}\) and \(P\cdot Q \,g^{\mu\nu}\) which are \(\mathcal{O}(g)\). For the retarded self-energy all terms with  \(P^{\mu}P^{\nu}\) cancel giving a suppression in \(g\). As an example we can look at Fig. \ref{fig:quarkloop_ra}. The diagram goes like
\begin{align}
\begin{split}
&\int d^4 P\; \frac{P^{\mu} P^{\nu} \delta(P^2)}{(Q+P)^2 + i\epsilon (q^0 + p^0)} \\
&\sim \int d^3 p\; \frac{P^{\mu} P^{\nu}}{2 q^0 p^0 - 2\mathbf{p}\cdot \mathbf{q} + i \epsilon p^0} \bigg|_{p^0 = p} \\
&+ \int d^3 p\; \frac{P^{\mu} P^{\nu}}{2 q^0 p^0 - 2\mathbf{p}\cdot \mathbf{q} + i \epsilon p^0} \bigg|_{p^0 = -p}
\end{split}
\end{align}
at leading order. The two terms cancel as can be seen by doing \(\mathbf{p} \rightarrow -\mathbf{p}\) in the last integral. Such a cancellation does not take place when the vertex factor is  \(P^{\mu}Q^{\nu} + Q^{\mu}P^{\nu}\) or \(P\cdot Q\, g^{\mu\nu}\).

The gluon \(rr\) propagator is more complicated in a non-equilibrium plasma than in equilibrium. In particular it has an imaginary part. Using the definition
\beq 
G_{rr}^{\mu\nu}(P) = \frac{1}{2}  \int d^4 x \; e^{iP \cdot x} \left\langle \left\{A^{\mu}(x), A^{\nu}(0)\right\} \right\rangle
\eeq
we get that \(G_{rr}^{\mu\nu}(P)^{*} =  G_{rr}^{\mu\nu}(-P)\). In translationally invariant systems we furthermore get that 
\begin{align} \label{eq:G_rr_symmetry}
\begin{split}
G_{rr}^{\mu\nu}(P) &= \frac{1}{2} \int d^4 x \; e^{-iP \cdot x}  \left\langle \left\{A^{\mu}(-x), A^{\nu}(0)\right\} \right\rangle \\
&= \frac{1}{2}  \int d^4 x \; e^{-iP \cdot x} \left\langle \left\{A^{\nu}(x), A^{\mu}(0)\right\} \right\rangle \\
&= G_{rr}^{\nu\mu}(-P)
\end{split}
\end{align} 
where we did a change of variables \(x \rightarrow -x\) in the first line and translated the propagator in the second line.
These results are clearer when writing  
\beq 
G_{rr}^{\mu\nu}(P) = \RE G_{rr}^{\mu\nu}(P) + i \,\IM G_{rr}^{\mu\nu}(P).
\eeq 
Then \(\RE G_{rr}^{\mu\nu}(-P) = \RE G_{rr}^{\mu\nu}(P)\) and \(\IM G_{rr}^{\mu\nu}(-P) = -\IM G_{rr}^{\mu\nu}(P)\). Furthermore,  Eq.\eqref{eq:G_rr_symmetry} shows that \(\RE G_{rr}\) is symmetric and \(\IM G_{rr}\) is antisymmetric under the interchange of the spacetime indices. Evaluation of \(G_{rr}\) explicitly given some momentum distribution function \(f\) is a subject for future research. The \(00\) component has been evaluated using an anisotropic momentum distribution as it has applications to the heavy-quark potential in QGP \cite{Nopoush2017}.

The imaginary part of \(G_{rr}\) might seem surprising. It is helpful to consider how it comes about. We can always write 
\beq 
G_{rr} = G_{\ret} \frac{-i\left(\Pi_{>} +\Pi_{<}\right) }{2}G_{\adv}.
\eeq
where 
\beq 
\left(1-G_{\ret}^0 \left(-i\Pi_{\ret}\right) \right) G_{\ret} = G_{\ret}^0
\eeq
and similarly for \(G_{\adv}\). 
The bare propagator and the HTL self energy is symmetric in the spacetime indices  so the same goes for 
\(G_{\ret}\) and 
\(G_{\adv}\). 
Thus \(G_{rr}\) is the product of three symmetric matrices. In general
\beq 
G_{rr}^{T} = G_{\adv} \frac{-i\left(\Pi_{>} +\Pi_{<}\right) }{2} G_{\ret}
\eeq  
will be different from \(G_{rr}\) because these matrices do not commute. In equilibrium (and for any isotropic momentum distribution) the only available tensors are \(g^{\mu\nu}\), the external momentum \(P^{\mu}\) and the plasma four-velocity, \(u^{\mu}\). Therefore all matrices are spanned by \(g^{\mu\nu}\), \(P^{\mu}P^{\nu}\) and the projection operators \(P_T\) and \(P_L\) \cite{Kapusta2006}. These four matrices commute so  \(G_{rr}^T = G_{rr}\) and \(G_{rr}\) is real. In an anisotropic plasma there are additional tensors describing the anisotropy and therefore more matrices. They will not all commute in general giving \(G_{rr}\) an imaginary part.

\section{The occupation density of hard quarks} \label{Hard_quarks}

To evaluate the LPM effect we need the \(rr\) propagator for hard and nearly on-shell quarks, i.e. \(S_{rr}(P)\) with \(P \sim T \) and \(P^2 \sim g^2 T^2\). We will show that at leading order
\begin{equation} \label{eq:S_rr_heuristic}
S_{rr} = \left[ \frac{1}{2} - F \right] \left(S_{\ret} - S_{\adv} \right),
\end{equation}
just as for scalars. Here \(F(P) := - P\cdot\Sigma_</2i P\cdot \IM \Sigma_{\ret}\) is a resummed occupation density.

%\begin{subfigures} 
%\centering
%	\begin{figure}{0.192\textwidth}
%	\includegraphics[width=\textwidth]{Fig6a.pdf} 
%	\caption{} 
%	\end{figure}
%	
%	\begin{figure}{0.192\textwidth}
%	\includegraphics[width=\textwidth]{Fig6b.pdf}
%	\caption{}
%	\end{figure}
%	\caption{Diagrams contributing to \(\Sigma_{\ret}\) at leading order in \(g\).}
%	\label{fig:Sigma_ret}
%\end{subfigures}

\begin{figure} 
\centering
	\begin{minipage}{0.32\textwidth}
	\includegraphics[width=0.6\textwidth]{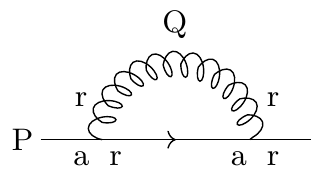} \\ 
	(a)\\ \vspace*{0.1cm}
	\end{minipage}
	\quad
	\begin{minipage}{0.32\textwidth}
	\includegraphics[width=0.6\textwidth]{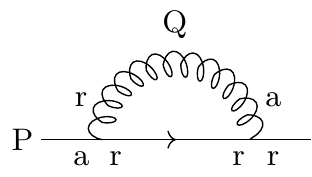}\\
        (b)
	\end{minipage}
	\caption{Diagrams contributing to \(\Sigma_{\ret}\) at leading order in \(g\).}
	\label{fig:Sigma_ret}
\end{figure}

We begin by evaluating \(\Sigma_{\ret}\). The contributing diagrams can be seen in Fig. \ref{fig:Sigma_ret}. For an internal particle with soft momentum, \(\mathcal{O}(gT)\), we must use a HTL resummed propagator, while for hard particles  we use bare propagators. There are a few different momentum regimes. When the loop momentum is hard, \(Q \sim T\), the two diagrams give rise to the thermal mass  
\begin{align} \label{eq:quark_thermal_mass}
\begin{split} 
m_{\infty}^2 &= 2P \cdot \RE \Sigma_{\ret}(P) \\
&= 2g^2 C_F \int \frac{d^3 p}{(2\pi)^3} \; \frac{2f_g(\mathbf{p}) +  f_q(\mathbf{p})+f_{\bar{q}}(\mathbf{p})}{2p}. 
\end{split}
\end{align}
The contribution of this momentum regime to \(\IM \Sigma_{\ret}\) is phase space suppressed because both \(G^0_{rr}\)
 and \(\IM S^0_{\ret}\) contain a delta function forcing the internal particles to be on shell \footnote{In our notation \(\Sigma_{\ret}\) can denote both a spinor matrix and a four-vector, i.e. \(\Sigma_{\ret} = \Sigma_{\ret}^{\mu} \gamma_{\mu}\).}.

We now focus on the top diagram in Fig.  \ref{fig:Sigma_ret} which is given by 
\begin{equation} 
\Sigma_{\ret} (P) \Big\rvert_{(a)} = -i g^2 C_F\int \frac{d^4 Q}{(2\pi)^4}\,  G_{rr}^{\mu\nu}(Q)\; \gamma_{\mu} S_{\ret}(P-Q)\gamma_{\nu}. 
\end{equation}
When the gluon is soft and the quark is hard the leading order contribution is
\begin{eqnarray} 
\Sigma_{\ret}(P) \Big\rvert_{\mathrm{soft}} & =   g^2 C_F \int \frac{d^4 Q}{(2\pi)^4}\, \gamma_{\mu} \slashed{P} \gamma_{\nu}\,
G^{\mu\nu}_{rr}(Q)\nonumber \\
& \times \left[  - i\pi \,\mathrm{sgn}(p^0)\, \delta(2P\cdot Q) + \frac{1}{-2P\cdot Q}\right] 
\end{eqnarray}
where we have substituted the bare quark propagator. We have used that the quark is on-shell, \(P^2 \sim g^2 T^2\).
Using the properties of \(G^{\mu\nu}_{rr}(Q)\) under the interchange of the spacetime indices and under \(Q \rightarrow -Q\) we can write
\beq \label{eq:soft_sigma_ret}
\Sigma_{\ret}(P) \Big\rvert_{\mathrm{soft}} = \RE \Sigma_{\mu}^{\mathrm{p}} \,\gamma^{5} \gamma^{\mu} + i\, \IM \Sigma_{\mu}\, \gamma^{\mu}
\eeq
where
\begin{eqnarray}
\IM \Sigma_{\omega} &=   - g^2 C_F\int^{gT} \frac{d^4 Q}{(2\pi)^4}\, \left(P_{\mu} g_{\nu\omega} + P_{\nu} g_{\mu\omega} - g_{\mu\nu} P_{\omega} \right) \nonumber \\
& \times \pi \mathrm{sgn}(p^0)\, \delta(2P\cdot Q)\, \RE G_{rr}^{\mu\nu} 
\end{eqnarray}
and we have a pseudovector component 
\beq
\RE \Sigma_{\omega}^{\mathrm{p}} = g^2 C_F\, \varepsilon_{\omega\rho\mu\nu} P^{\rho} \,\int^{gT} \frac{d^4 Q}{(2\pi)^4}\,   \frac{\IM G_{rr}^{\mu\nu}}{2P\cdot Q}.
\eeq
Here we used the identity \cite{Pal2007}
\beq 
\gamma_{\mu} \gamma_{\rho} \gamma_{\nu} = g_{\mu\rho} \gamma_{\nu} + g_{\rho\nu} \gamma_{\mu} - g_{\mu\nu} \gamma_{\rho} + i \epsilon_{\sigma \mu \rho \nu} \gamma^{\sigma} \gamma^{5}
\eeq 
to separate the symmetrical and antisymmetrical part of \(G_{rr}\). 
The expression in Eq. \eqref{eq:soft_sigma_ret} is clearly leading order because \(G_{rr} \sim g^{-3}T^2\) and \(Q \sim gT\).  The vector term with \(\IM \Sigma_{\mu}\) is both present in equilibrium and non-equilibrium plasma. It determines the decay width of hard quarks, \(\Gamma\), through
\begin{align} \label{eq:quark_Gamma}
\begin{split}
-\frac{1}{2} p^0 \Gamma &= P \cdot \IM \Sigma_{\ret} \\
&=  -2 \pi g^2 C_F \; \mathrm{sgn}(p^0) \\
& \times P_{\mu} P_{\nu} \int^{gT}\frac{d^4 Q}{(2\pi)^4}\, \RE G_{rr}^{\mu\nu}(Q)\;  \delta(2P\cdot Q).
\end{split}
\end{align} 
 The pseudovector term with \(\RE \Sigma^p_{\mu}\) is only present in anisotropic systems because it includes the imaginary part of \(G_{rr}\).

There are no further leading order contributions to \(\Sigma_{\ret}\). The case of a soft internal quark, \(P-Q \sim gT\), in part (a) of Fig. \ref{fig:Sigma_ret} is subleading because there is no enhancement from soft gluons. Similarly, diagram (b) of the same figure is subleading when either particle is soft because the \(g^{-3}\) contribution from \(G_{rr}\) is missing.

We can now derive the retarded quark propagator. We have shown that the full self-energy is 
\beq 
\Sigma_{\ret} = \Sigma_{\mu} \,\gamma^{\mu} + \RE \Sigma^{\mathrm{p}}_{\mu}\, \gamma^{5} \gamma^{\mu} 
\eeq
where \(\RE \Sigma_{\mu}\) gives the thermal mass in Eq. \eqref{eq:quark_thermal_mass} and \(\IM \Sigma_{\mu}\) and \(\RE \Sigma^{\mathrm{p}}_{\mu}\) are as before. The retarded propagator in the Dirac representation is then
\beq
S_{\ret} =
\begin{bmatrix}
0 & \frac{i\left(P-\Sigma + \RE \Sigma^{\mathrm{p}} \right)\cdot \sigma}{\left(P-\Sigma + \RE \Sigma^{\mathrm{p}} \right)^2} \\
\frac{i\left(P-\Sigma - \RE \Sigma^{\mathrm{p}} \right)\cdot \bar{\sigma}}{\left(P-\Sigma - \RE \Sigma^{\mathrm{p}} \right)^2} & 0
\end{bmatrix}
\eeq  
Because of the \(\gamma^5\) matrix the  pseudovector part of the self-energy, \(\RE \Sigma^{\mathrm{p}}\), has different signs for positive and negative helicities. However, both helicities have the same thermal mass since \(P \cdot \RE \Sigma^p_{\ret}\) vanishes because of the Levi-Civita tensor. The imaginary part of the self energy still has the same sign for both helicities.  
Thus we can write  
\beq \label{eq:S_ret}
S_{\ret} = \frac{i \slashed{P}}{P^2 - m_{\infty}^2 + i \Gamma p^0}.
\eeq
at leading order where \(m_{\infty}^2\) is given by Eq. \eqref{eq:quark_thermal_mass}. We have used that \(\left(P - \Sigma \right)^2 \approx P^2 - 2P\cdot \Sigma\). 
We see that the pseudovector component, and therefore \(\IM G_{rr}\), does not contribute when considering on-shell particles  which simplifies the calculations considerably.

We can finally derive \(S_{rr}\) in Eq. \eqref{eq:S_rr_heuristic}. A similar argument as for \(\Sigma_{\ret}\) shows that 
\begin{align}
\begin{split}
P \cdot \Sigma_{<} (P) = &\;4\pi i g^2 C_F \left[f_q(\mathbf{p}) \theta(p^0) + (f_{\bar{q}}(-\mathbf{p})-1) \theta(-p^0) \right]\\
&\times  P_{\mu} P_{\nu}  \int^{gT} \frac{d^4 Q}{(2\pi)^4}\, \RE G_{rr}^{\mu\nu}(Q)\; \delta(2P\cdot Q). 
\end{split}
\end{align}
The quark momentum distribution comes from \(S_{12}^0\) and we have used that \(G_{12} \approx G_{rr}\) for soft gluons. 
Since
\beq
\slashed{P} \slashed{\Sigma}_< \slashed{P} = - P^2\, \slashed{\Sigma}_< + 2P \cdot \Sigma_< \,\slashed{P}
 \approx 2P \cdot \Sigma_<\, \slashed{P}
\eeq
one can easily show that at leading order  
\begin{align}
\begin{split}
S_{rr}(P) &= \frac{1}{2}  \left(S_{\ret} - S_{\adv} \right) + S_{\ret} \left(-i \Sigma_< \right) S_{\adv}  \\
&\approx \left( \frac{1}{2} - F(P) \right) \left(S_{\ret} - S_{\adv}  \right). 
\end{split}
\end{align} 
where
\begin{equation} \label{eq:quarkoccupation}
F(P) := - \frac{P \cdot \Sigma_<}{2iP \cdot \IM \Sigma_{\ret}} =  f_q(\mathbf{p})\, \theta(p^0) + (1-f_{\bar{q}}(-\mathbf{p}))\, \theta(-p^0).
\end{equation}
All dependence on \(G_{rr}\) cancels out in \(F\). 

Comparing with the expression for \(S_{rr}\) in equilibrium we see that \(F\) should be interpreted as a resummed occupation density. Its form makes perfect sense. 
In the Boltzmann equation incoming particles have \(p^0>0\) and outgoing particles have \(p^0 < 0\). 
Thus \(F\) is just the bare momentum distribution with Pauli blocking for outgoing quarks.
This function reduces to the Fermi-Dirac distribution \(f_{F}(p^0)\) in equilbrium as 
can be seen by using \(1-f_F(-x) = f_F(x)\) and noting that when going from equilibrium to non-equilibrium systems one makes the identification \(f_F(|p^0|) \leftrightarrow f_q(\mathbf{p})\).

%\begin{figure}[h]
%\centering
%
%	\begin{subfigure}{0.21\textwidth}
%	\includegraphics[width=\textwidth]{Fig7a.pdf} 
%	\caption{} 
%	\end{subfigure}
%	
%	\begin{subfigure}{0.28\textwidth}
%	\includegraphics[width=\textwidth]{Fig7b.pdf}
%	\caption{}
%	\end{subfigure}
%	
%	\caption{Definition of the four-point function \(S_{nmkl}(x_1,x_2;y_1,y_2)\) in position space. \(n\), \(m\), \(k\), \(l\) are either \(1\) or \(2\). Also shown is the diagram we need to evaluate, i.e. \(S_{1122}\) in momentum space.}
%	\label{fig:S_1122}
%\end{figure}

\begin{figure}[h]
\centering
	\begin{minipage}{0.3\textwidth}
	\includegraphics[width=0.7\textwidth]{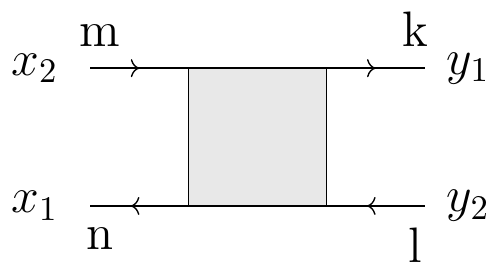}
	\end{minipage}\\
	(a)
	\hspace{2cm}
	\begin{minipage}{0.4\textwidth}
	\includegraphics[width=0.7\textwidth]{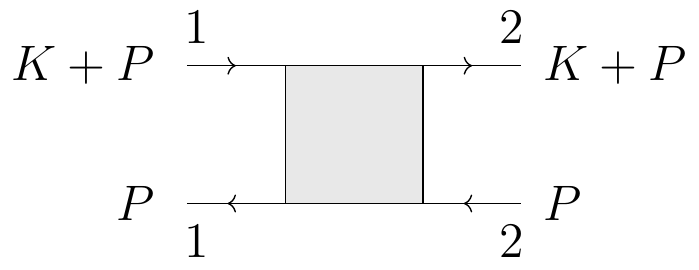}
	\end{minipage}\\
	(b)
	\caption{Definition of the four-point function \(S_{nmkl}(x_1,x_2;y_1,y_2)\) in position space. \(n\), \(m\), \(k\), \(l\) are either \(1\) or \(2\). Also shown is the diagram we need to evaluate, i.e. \(S_{1122}\) in momentum space.}
	\label{fig:S_1122}
\end{figure}

\section{The LPM effect in a non-equilibrium plasma} \label{Sum}

\begin{figure*}[th]
\begin{center}
\includegraphics[width=0.65\textwidth]{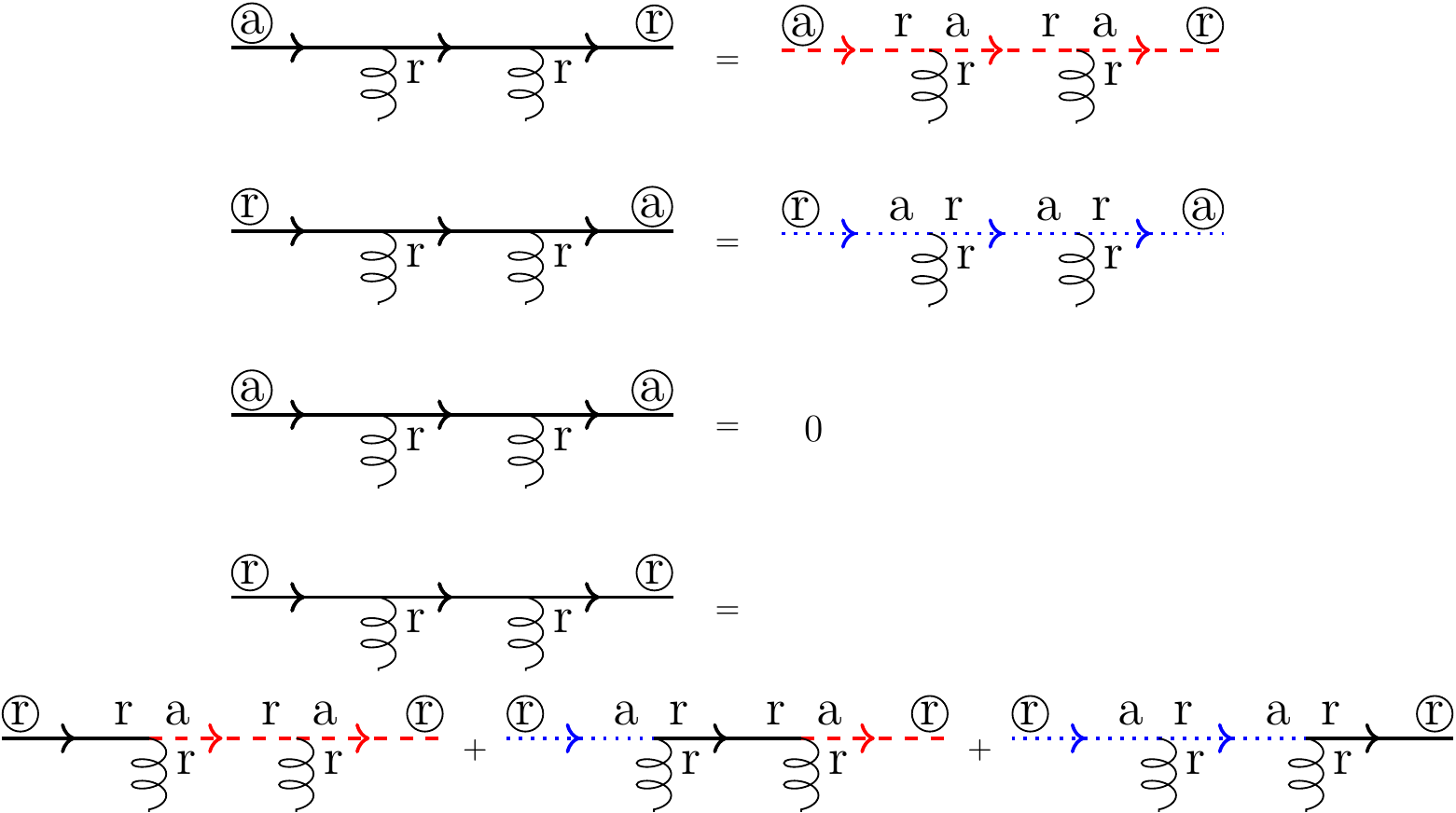}
\end{center}
\caption{(Color online) Analysis of quark rails in the \(r/a\) basis.}
\label{fig:rails}
\end{figure*}

We now have all the ingredients to evaluate the LPM effect in an out-of-equilibrium quark-gluon plasma, and to compute the photon production rate in Eq. (\ref{eq:photonrate}). 
The photon sources are connected to a quark and an antiquark so we need to evaluate the four-point function 
\beq 
S_{1122}(x_1,x_2;y_1,y_2) = \langle T_{\mathcal{C}} \left\{ \overline{\psi}_1(x_1) \psi_1(x_2) \overline{\psi}_2(y_1) \psi_2(y_2)\right\}\rangle. 
\eeq 
See Fig. \ref{fig:S_1122}, top diagram, for the corresponding contribution. When going to momentum space we can approximate the momentum in the quark or the antiquark rail as constant because it only changes through the exchange of soft gluons. The relevant diagram is the bottom one in Fig. \ref{fig:S_1122} where \(K\) is the photon momentum and \(P\) is the loop momentum. 

\subsection{Summing four-point functions without the KMS condition} 

\begin{figure}
\centering
	\begin{minipage}{0.2\textwidth}
	\includegraphics[width=0.7\textwidth]{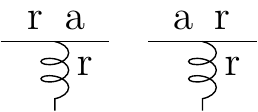}
	\end{minipage}
	\caption{(Color online) Vertices that contribute at leading order.}
	\label{fig:vertices}
\end{figure}

\begin{figure}
\centering
	\begin{minipage}{0.5\textwidth}
	\includegraphics[width=0.7\textwidth]{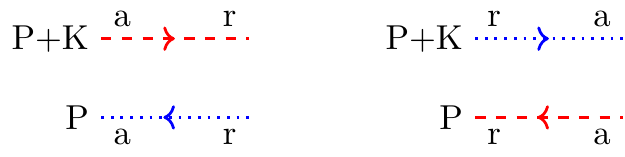}
	\end{minipage}
	\caption{(Color online) The pairs of propagators that give pinching poles.}
	\label{fig:pinch_poles}
\end{figure}

Up until now our analysis has been in the \(r/a\) basis which enables power counting of the complicated diagrams. We must evaluate \(S_{1122}\) using the expression  
\begin{align} \label{eq:S_1122}
\begin{split} 
S_{1122}& = S_{rrrr} + \frac{1}{2} \left(S_{arrr}  + S_{rarr} - S_{rrar} -  S_{rrra}\right) \\
 +& \frac{1}{4}\left( S_{aarr} - S_{arar} - S_{arra} - S_{raar}  - S_{rara}  + S_{rraa} \right)\\
 + & \frac{1}{8}\left( S_{raaa} + S_{araa} - S_{aara} - S_{aaar}  + \frac{1}{2} S_{aaaa}\right). 
\end{split}
\end{align} 
This task might look overwhelming. Each four-point function on the right hand side is a sum of infinitely many diagrams with a different number of gluon rungs. They need not have any clear pattern in their \(r/a\) indices. 

In thermal equilibrium the four-point functions have been related using the KMS condition \cite{Wang1998}. Specifically,
\begin{align} \label{eq:fourp_KMS}
\begin{split}
S_{1122} = & \alpha_1 S_{aarr} + \alpha_2 S_{aaar} + \alpha_3 S_{aara} + \alpha_4 S_{araa} \\
&+ \alpha_5 S_{raaa} +\alpha_6 S_{arra} +\alpha_7 S_{arar} + c.c.
\end{split}
\end{align}
where the coefficients depend on the Fermi-Dirac distribution. As an example \(\alpha_1 = f_F(p^0 + k^0)\left(1-f_F(p^0) \right)\). One can also show  that \(S_{aarr}\) is the only one of these four-point functions that contributes at leading order. Therefore
\beq \label{eq:S1122_KMS}
S_{1122} = 2 f_F(p^0 + k^0)\left(1-f_F(p^0) \right) \RE S_{aarr}
\eeq
in thermal equilibrium.

\begin{figure}
\centering
	\includegraphics[width=0.21\textwidth]{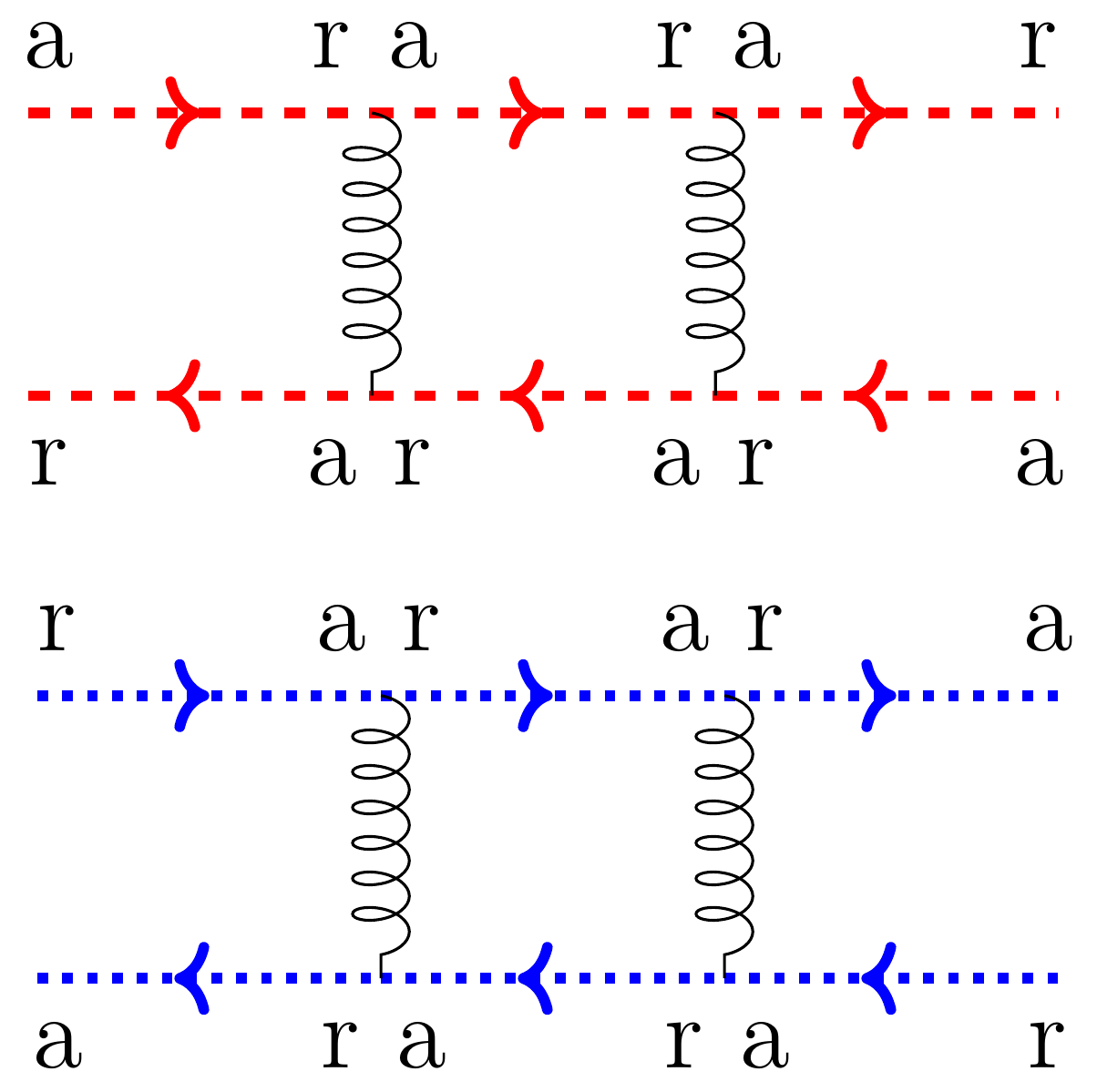}
	\caption{ (Color online) \(S_{rara}\) and \(S_{arar}\). These diagrams do not contribute at leading order because there are no pinching poles.}
	\label{fig:rara_arar}
\end{figure}

\begin{figure}
\centering
	\begin{minipage}{0.4\textwidth}
	\includegraphics[width=0.7\textwidth]{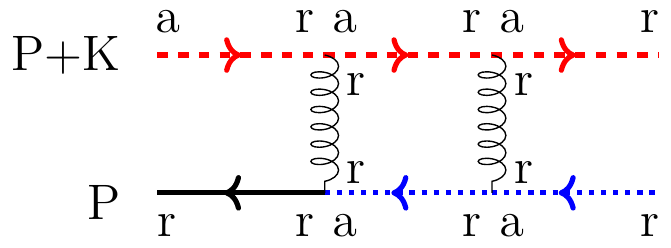}
	\end{minipage}
	\caption{(Color online) The only way of placing \(r/a\) indices in \(S_{rarr}\) at leading order. In a general diagram with arbitrarily many gluon rungs \(S_{rr}\) must still be on the far left.}
	\label{fig:S_rarr}
\end{figure}

\begin{figure}
\centering
	\begin{minipage}{0.4\textwidth}
	\includegraphics[width=0.7\textwidth]{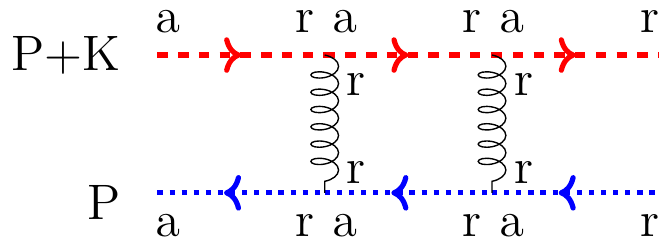}
	\end{minipage}
	\caption{(Color online) The only way of placing \(r/a\) indices in \(S_{aarr}\) at leading order.}
	\label{fig:S_aarr}
\end{figure}

We will now evaluate \(S_{1122}\) generally without using the KMS condition. Our derivation is thus also valid in non-equilibrium systems. It only relies on the power counting scheme. We know that all gluon rungs must be \(rr\) propagators to get the \(1/g\) enhancement from the density of soft gluons. Each vertex contains an odd number of \(a\) indices so one quark propagator ends with \(a\) and one with \(r\) at each vertex, see Fig. \ref{fig:vertices}. Finally \(aa\) propagators vanish.

We first consider quark rails as in Fig. \ref{fig:rails}. For the convenience of the reader we draw \(S_{ar}\) with red, dashed lines, \(S_{ra}\) with blue, dotted lines and \(S_{rr}\) with black lines. The first propagator in a quark rail that starts with \(a\) must be \(S_{ar}\). The next propagator must then start with \(a\) and so on. Thus 
all propagators are \(S_{ar}\) at leading order. Similarly all propagators in a quark rail that starts with \(r\) and ends with \(a\) are \(S_{ra}\). Finally, quark rails that start and end with \(a\) vanish. This means that we can ignore \(S_{araa}\), \(S_{aara}\), \(S_{arra}\), \(S_{raar}\), \(S_{aaar}\), \(S_{raaa}\) and \(S_{aaaa}\) at leading order. (In fact, \(S_{aaaa}\) vanishes at all orders \cite{Wang1998}.)

There are more possibilities for quark rails that start and end with \(r\). It is easy to see that they consist of arbitrarily many \(S_{ra}\), then one \(S_{rr}\), and finally arbitrarily many \(S_{ar}\). Thus there are \(n\) possibilities for a quark rail with \(n\) propagators which differ in the placement of \(S_{rr}\). See Fig. \ref{fig:rails} for the case of three propagators.

The remaining four-point functions can only contribute at leading order if we get pinching poles from each pair of propagators between adjacent gluon rungs.  
Just like in Eq. \eqref{eq:pinching_poles} that means that one propagator is \(S_{ra}\) and the other one \(S_{ar}\), see Fig. \ref{fig:pinch_poles}. This immediately tells us that \(S_{rara}\) and \(S_{arar}\) can be discarded because they have no pinching poles, see Fig. \ref{fig:rara_arar}. 

\begin{figure}[b!]
\centering
	\begin{minipage}{0.3\textwidth}
	\includegraphics[width=0.7\textwidth]{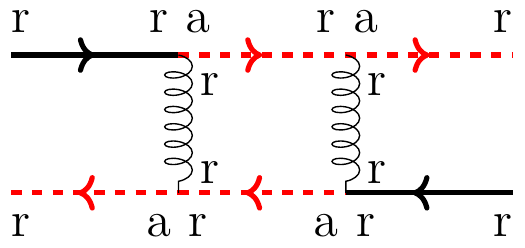}
	\end{minipage}
	\caption{(Color online) An example of a diagram that does not contribute to \(S_{rrrr}\) at leading order because the propagators in the middle do not give pinching poles.}
	\label{fig:rrrr_no_pinching}
\end{figure}

\begin{figure*}[t!]
\begin{center}
\includegraphics[width=0.83\linewidth]{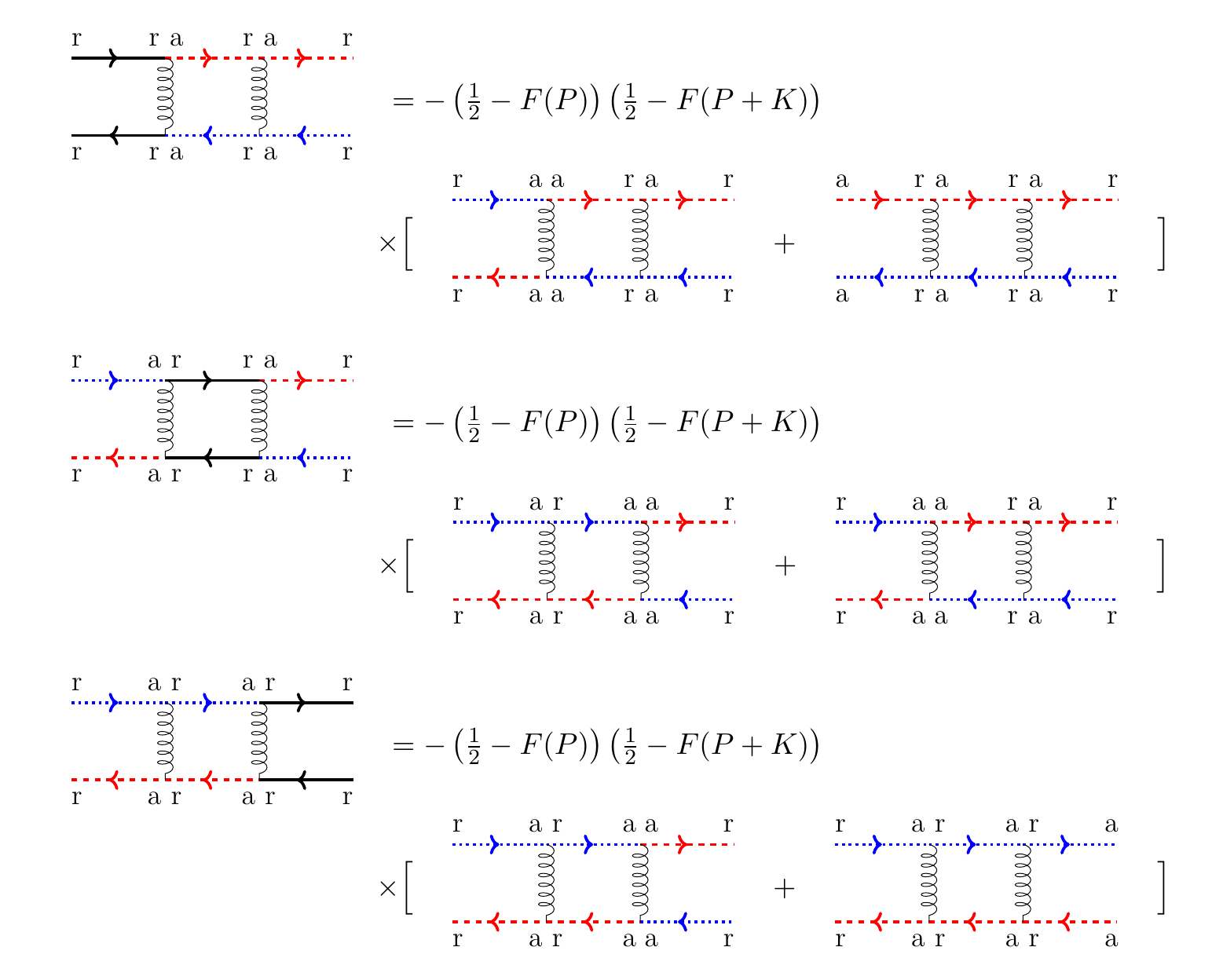}
\end{center}
\caption{(Color online) Leading order contributions to \(S_{rrrr}\) where the \(S_{rr}\) propagators are on top of each other.}
\label{fig:rrrr_top}
\end{figure*}

\begin{figure*}[ht]
\begin{center}
\includegraphics[width=0.78\linewidth]{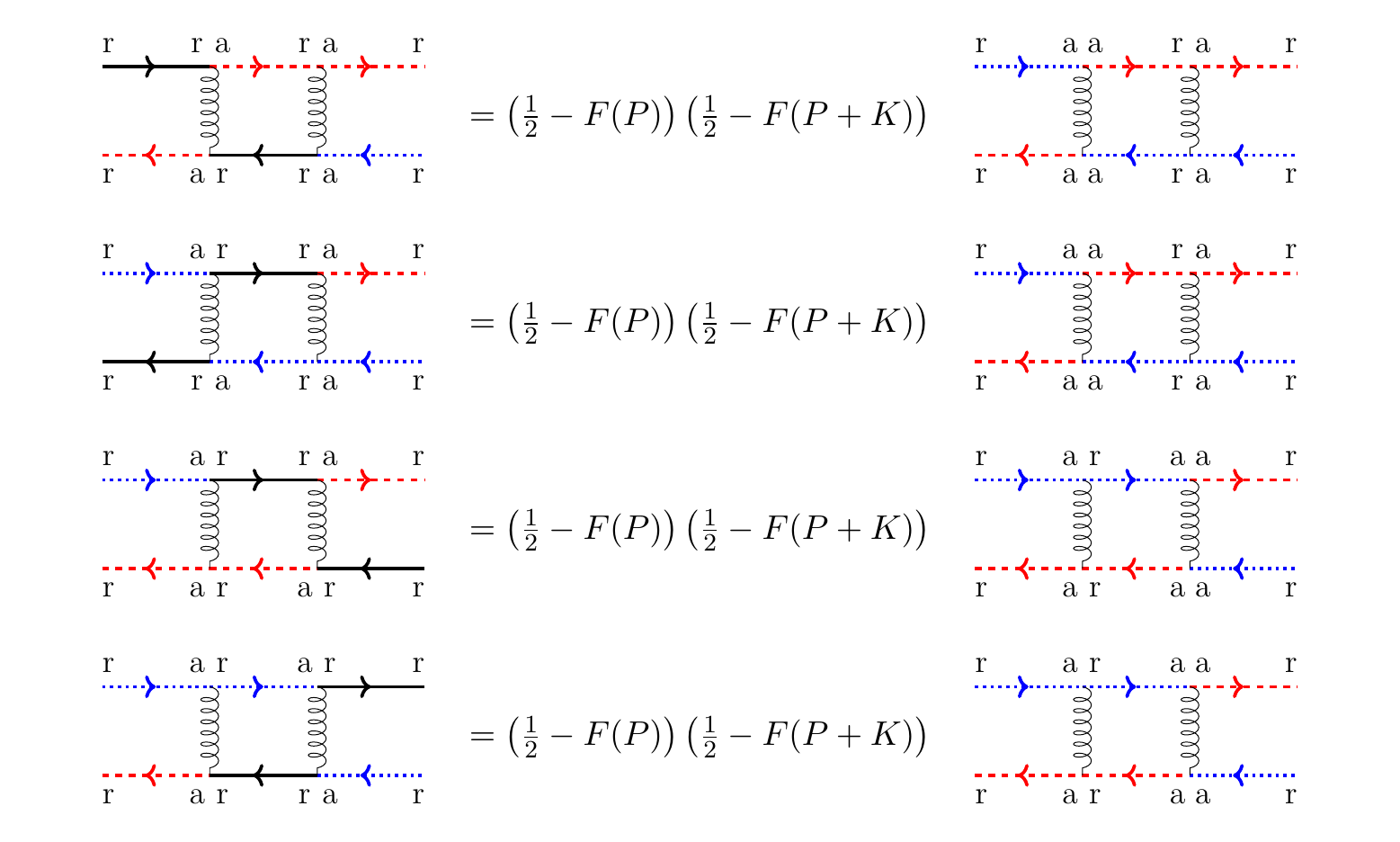}
\end{center}
\caption{(Color online) Leading order contributions to \(S_{rrrr}\) where the \(S_{rr}\) propagators are immediately diagonal to each other.}
\label{fig:rrrr_diag}
\end{figure*}

\begin{figure*}[ht!]
\begin{center}
\includegraphics[width=0.78\linewidth]{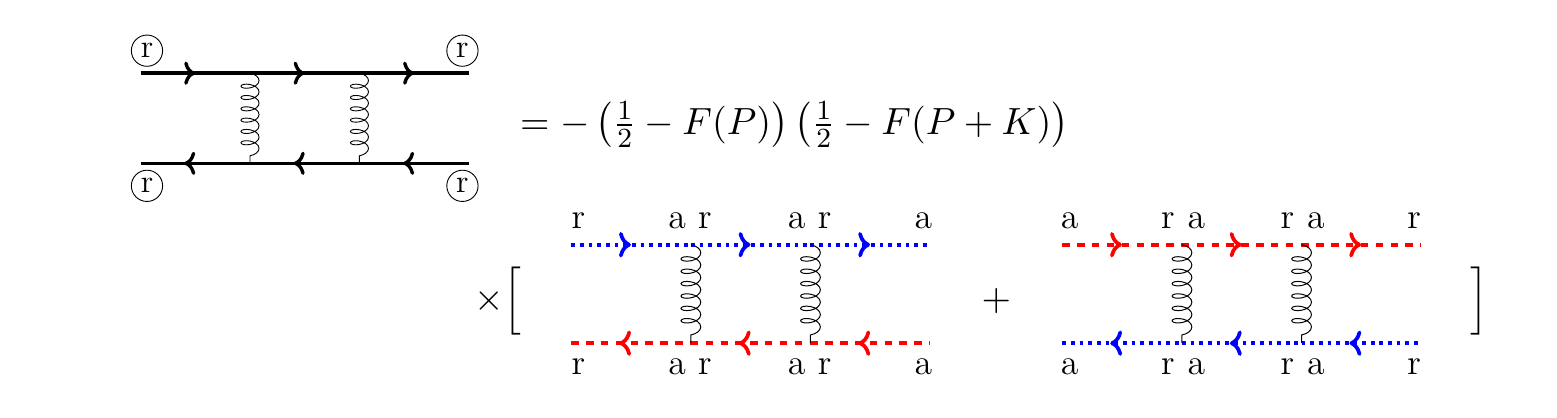}
\end{center}
\caption{(Color online) All leading order contributions to \(S_{rrrr}\) after cancellation.}
\label{fig:rrrr_remainder}
\end{figure*}

We can now express the remaining seven four-point functions in terms of \(S_{aarr}\) and \(S_{rraa}\). They all include an \(rr\) propagator the pole structure of which can be seen from  
\beq \label{eq:S_rr}
S_{rr} = \left[ \frac{1}{2} - F(P) \right] \left( S_{ra} - S_{ar} \right).
\eeq  
It has poles on both sides of the real axis. At leading order we can drop the term that does not give a pinching pole. As an example we must place the indices in \(S_{rarr}\) as in Fig. \ref{fig:S_rarr} to get pinching poles from all pairs. Then \(S_{rr}\) is on the far left and only the term with \(S_{ra}\) contributes. Thus
\beq  \label{eq:S_rarr}
S_{rarr} = \left( \frac{1}{2} - F(P)\right) S_{aarr}.
\eeq 
where the leading order diagram for \(S_{aarr}\) is in Fig. \ref{fig:S_aarr}.
Similarly, one sees that 
\begin{align}
S_{rrar} &= - \left( \frac{1}{2} - F(P)\right) S_{rraa} \\
S_{arrr} &= - \left( \frac{1}{2} - F(P+K)\right) S_{aarr} \\
S_{rrra} &= \left( \frac{1}{2} - F(P+K)\right) S_{rraa}.
\end{align}

To finish our derivation, we analyze \(S_{rrrr}\) which is more complicated. Each quark rail has one \(rr\) propagator. The two \(S_{rr}\) must be on top of each other or immediately diagonal to each other since otherwise we will miss a pinching pole pair, see Fig. \ref{fig:rrrr_no_pinching}. Let's consider the case of two gluon rungs. In Fig. \ref{fig:rrrr_top} we analyze the three possibilities of having the \(rr\) on top of each other. The remaining possibilities are analyzed in Fig. \ref{fig:rrrr_diag}. When these contributions are summed over, all terms cancel except for those corresponding to \(S_{aarr}\) and \(S_{rraa}\), see Fig. \ref{fig:rrrr_remainder}. A similar cancellation takes place for any number of gluon rungs. Specifically, four-point functions with \(S_{rr}\) immediately diagonal to each other cancel out with four-point functions with \(S_{rr}\) on top of each other. We are then left with
\beq  \label{eq:S_rrrr}
S_{rrrr} = - \left(\frac{1}{2} - F(P) \right) \left( \frac{1}{2} - F(P+K)\right) \left[ S_{rraa} + S_{aarr}\right].
\eeq

\begin{figure*}
\begin{center}
\includegraphics[width=0.86\linewidth]{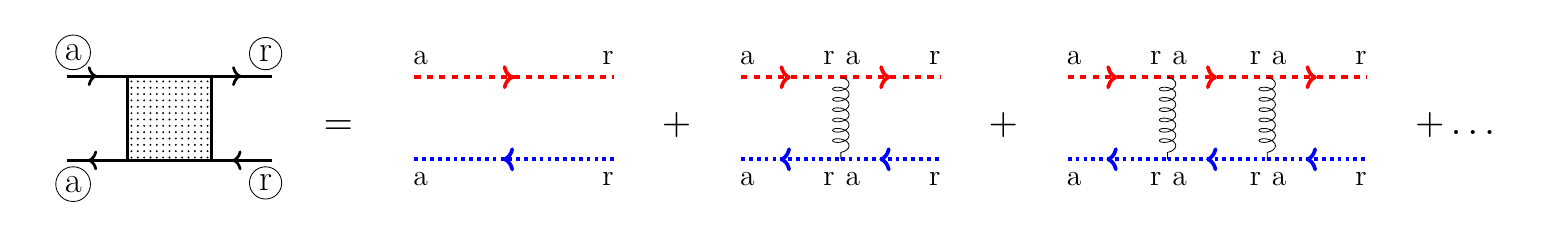}
\end{center}
\caption{(Color online) Leading order diagrams contributing to \(S_{aarr}\).}
\label{fig:rraa}
\end{figure*}

We have evaluated all terms in Eq. \eqref{eq:S_1122} at leading order. Summing up Eq. \eqref{eq:S_rarr} to \eqref{eq:S_rrrr} we get that
\beq 
S_{1122} = F(P+K) \left( 1- F(P)\right) \left[ S_{rraa} + S_{aarr}\right].
\eeq
This can be rewritten using \(S_{rraa} = S_{aarr}^*\) which can be seen from the definition of the four-point functions or from \(S_{ra}^{*} = -S_{ar}\). We have thus shown that
\beq \label{eq:1122andaarr}
S_{1122} = 2 F(P+K) \left( 1- F(P)\right) \RE S_{aarr}
\eeq
without using the KMS condition. This expression is convenient because \(S_{aarr}\) has a very simple structure, see Fig. \ref{fig:rraa}. In thermal equilibrium \(F\) from Eq. \eqref{eq:quarkoccupation} reduces to the Fermi-Dirac distribution and we retrieve the equilibrium result, Eq. \eqref{eq:S1122_KMS}. 

The factors of \(F\) in Eq. \eqref{eq:1122andaarr} give the momentum distribution of incoming and outgoing quarks including Pauli blocking. Our analysis is valid for on-shell photons, \(k^0 \approx k\). The momentum regime where \(p^0 > 0\) represents bremsstrahlung off a quark with initial momentum \(\mathbf{k} + \mathbf{p}\). The distribution functions are \(f_q(\mathbf{k} + \mathbf{p}) \left(1-f_q(\mathbf{p}) \right)\) as expected because there is one incoming and one outgoing quark. Bremsstrahlung off an antiquark is given by \(p^0 < -k\). The antiquark's initial momentum in the photon's direction is \(-p^0\) and the final momentum is \(-(k+p^0)\). The distribution functions are then \(\left( 1-f_{\bar{q}}(-\mathbf{k} - \mathbf{p})\right) f_{\bar{q}}(-\mathbf{p})\). The antiquark distributions are evaluated at negative momentum because we defined the momentum to flow in the direction of quarks. Finally the momentum regime \(-k < p^0 < 0\) corresponds to the pair annihilation of a quark with momentum \(k + p^0\) and an antiquark with momentum \(-p^0\). The distribution functions are \(f_q(\mathbf{k} + \mathbf{p}) f_{\bar{q}}(-\mathbf{p})\).

\subsection{Summing ladder diagrams}

\begin{figure}
\begin{center}
\includegraphics[width=\columnwidth]{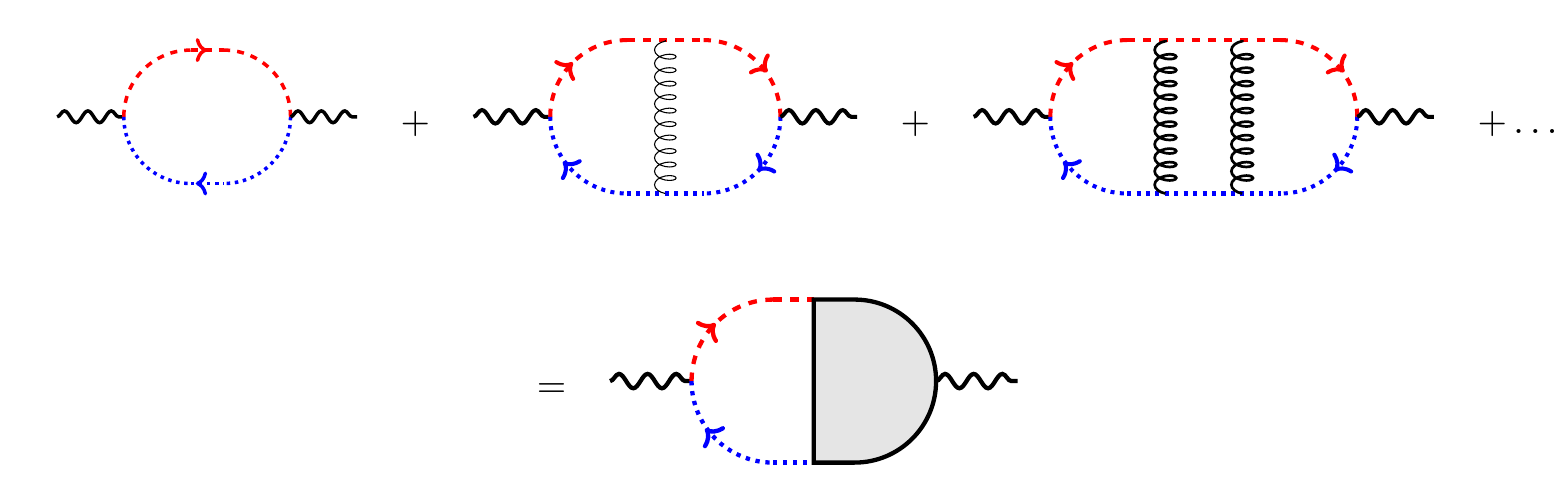}
\end{center}
\caption{(Color online) The LPM diagrams that contribute at leading order to the photon polarization tensor \(\Pi^{\gamma}_{12}\). Red propagators are \(S_{ar}\) and blue propagators are \(S_{ra}\).}
\label{fig:laddersum}
\end{figure}

\begin{figure}
\begin{center}
\includegraphics[width=0.65\columnwidth]{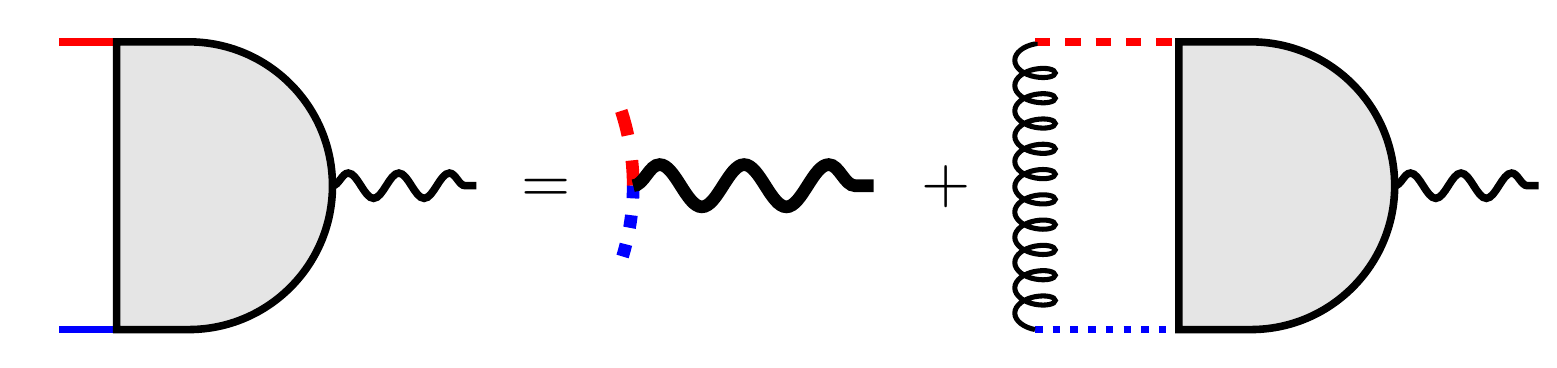}
\end{center}
\caption{(Color online) Procedure for summing up the diagrams in Fig. \ref{fig:laddersum}.}
\label{fig:Bethe_Salpeter}
\end{figure}

\begin{figure}
\begin{center}
\includegraphics[width=0.35\columnwidth]{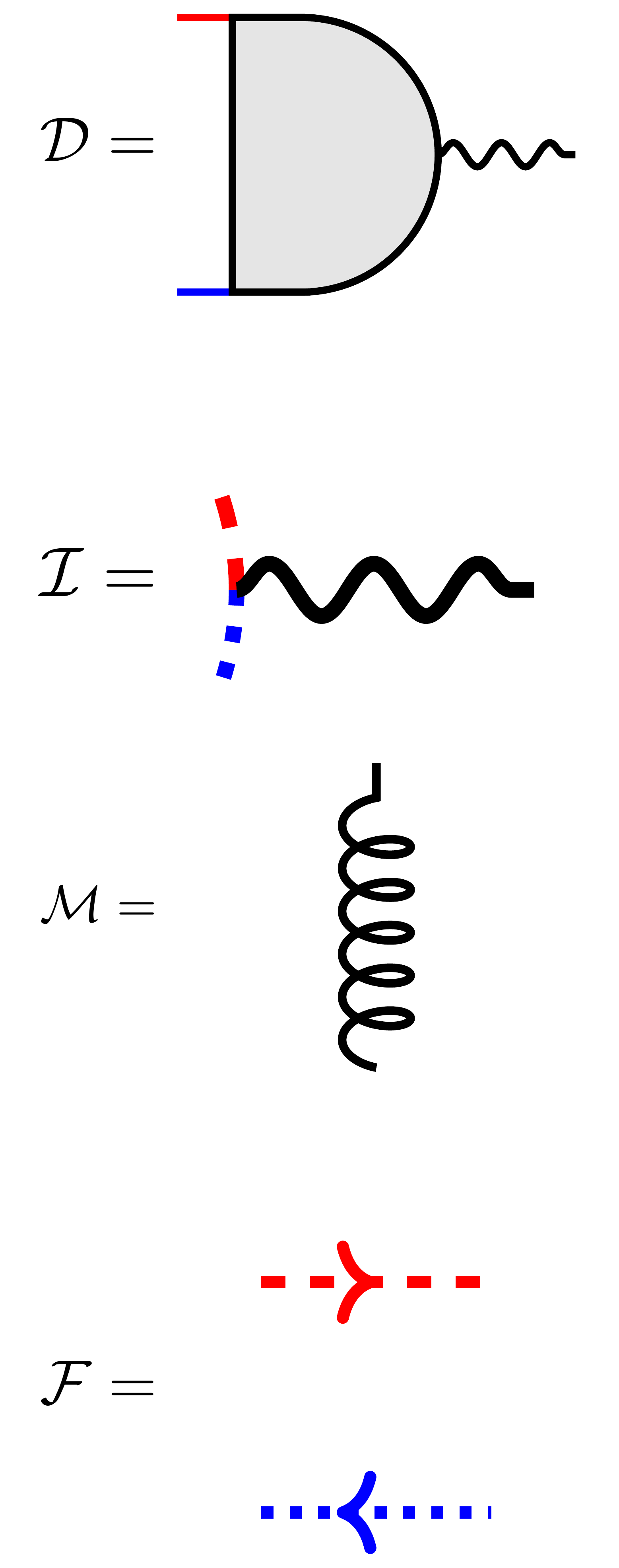}
\end{center}
\caption{(Color online) Definition of the quantities in Eq. (\ref{eq:formalBethe}).}
\label{fig:def_Bethe}
\end{figure}

The only remaining task is to sum up the ladder diagrams contributing to \(S_{aarr}\). This gives an integral equation which describes the LPM effect. We will outline the derivation briefly as it is quite similar to the one in thermal equilibrium, see \cite{Arnold2001,Mamo2015} for further details. The notation follows that of Ref. \cite{Jeon1994}. The contributing diagrams are shown in Fig. \ref{fig:laddersum} and the procedure for summing them up is in Fig. \ref{fig:Bethe_Salpeter} and \ref{fig:def_Bethe}. One gets an integral equation for the resummed vertex \(\mathcal{D}(P,P+K)\) which can be written schematically as
\beq \label{eq:formalBethe}
\mathcal{D}^{\mu} = \mathcal{I}^{\mu} + \int\hspace{-0.15cm}\frac{d^4Q}{\left(2\pi\right)^4} \; \mathcal{M} \mathcal{F} \mathcal{D}^{\mu}. 
\eeq 
Here \(\mathcal{I}^{\mu}\) is the bare vertex, \(\mathcal{F}(P,P+K)\) are quark propagators that give pinching poles and \(\mathcal{M}(P,P+K)\) is the contribution from a gluon ladder. The gluon momentum \(Q\) is soft, \(K\) is the photon momentum and \(P\) is the momentum in one of the quark rails. We need to evaluate \(\mathcal{F}\) and \(\mathcal{M}\).

We let the photon propagate in the z-direction. The collinear momentum of the quark momentum is hard, \(p^z \sim T\), but the orthogonal components are soft, \(\mathbf{p}_{\perp} \sim gT\). Furthermore the quark is nearly on shell, \(p^0 = p^z + \mathcal{O}(g^2T)\). It is convenient to decompose the quark propagator in Eq. \eqref{eq:S_ret} by helicity, 
\beq
S_{\ret} =
\begin{bmatrix}
0 & S^L_{\ret} \\
S^R_{\ret} & 0
\end{bmatrix}
\eeq 
We will focus on the right-handed part. At leading order one can write it as 
\beq \label{eq:Sretpoles}
S^R_{\ret}(P) = \frac{i}{2p} \left[  \frac{vv^{\dagger}}{p^0 +E_{\mathbf{p}} + i \Gamma/2}  + \frac{u u^{\dagger}}{p^0  - E_{\mathbf{p}} + i \Gamma/2} \right]
\eeq 
where \(E_{\mathbf{p}} = \sqrt{p^2 + m_{\infty}^2}\) is the quasi-particle energy. This equation has the same form as in equilibrium but the thermal mass \(m_{\infty}^2\) and the decay width \(\Gamma\) are now out-of-equilibrium constants. We have defined \(u(\mathbf{p})\) and \(v(\mathbf{p})\) to be the eigenvectors of \(\bm{\sigma} \cdot \hat{\mathbf{p}}\) with positive and negative eigenvalues. Their normalization is \(v^{\dagger} v = u^{\dagger} u = 2p\) and they obey \(v v^{\dagger} = p - \bm{\sigma} \cdot \mathbf{p}\) and \(u u^{\dagger} = p + \bm{\sigma} \cdot \mathbf{p}\). The first term in Eq. \eqref{eq:Sretpoles} which has \(\RE p^0 < 0\) at the pole describes a left-handed antiquark while the second term describes a right-handed quark.  

For simplicity we consider the case when \(p^0 > 0\) so only the second term in Eq. \eqref{eq:Sretpoles} contributes. At each gluon vertex we get a factor
\beq 
u^{\dagger}(\mathbf{p}) \,\sigma^{\mu} \,u(\mathbf{p}) = 2 \,(p,\mathbf{p}) \approx 2 P^{\mu}.  
\eeq
where the spin indices \(u\) come from the adjacent propagators.
The gluon rungs then give 
\begin{align}
\begin{split}
\mathcal{M} = &\; - 4 g^2 C_F \,   P_{\mu} \left(K_{\nu}+ P_{\nu}\right)\, G^{\mu\nu}_{rr}(Q)\\
 \approx & \;- 4 g^2 C_F \, p^z (k+p^z)\, \hat{K}_{\mu} \hat{K}_{\nu}\, \RE G^{\mu\nu}_{rr}(Q).
\end{split}
\end{align}
The loop momentum \(P^{\mu}\) is nearly collinear with the photon momentum \(K^{\mu}\) so only the real and symmetric part of \(G_{rr}\) contributes.  Here \(\hat{K}^{\mu} = (1,0,0,1)\).

Using \(S^R_{\adv} = - S^{R\;*}_{\ret}\) the pinching pole contribution is easily evaluated to be 
\begin{align} 
\begin{split}
\int \frac{dp^0}{2\pi} \mathcal{F}(P;K) = &\int \frac{dp^0}{2\pi} S^R_{\adv}(K+P) S_{\ret}^R(P)\\ =&\;\frac{1}{4 p^z (p^z + k) \left[ \Gamma + i\delta E \right]}.
\end{split}
\end{align}  
where
\beq 
\delta E = k^0 + E_{\mathbf{p}} \;\mathrm{sgn}(p^z) - E_{\mathbf{p}+ \mathbf{k}} \;\mathrm{sgn}(p^z+k)
\eeq
is of order \(g^2\). When \(p^0 < 0\) one gets similar expressions.

We can now assemble all the pieces in Eq. \eqref{eq:formalBethe}. 
The quark decay width, Eq. \eqref{eq:quark_Gamma}, can be written as 
\beq \label{eq:decay_and_kernel}
\Gamma = \int \frac{d^2 q_{\perp}}{(2\pi)^2} \mathcal{C}(\mathbf{q}_{\perp})
\eeq 
with 
\begin{equation} \label{eq:coll_kern}
\mathcal{C}(\mathbf{q}_{\perp}) = g^2 C_F \int \frac{dq_0 dq_{z}}{(2\pi)^2} \;2\pi \delta(q_0 - q_{z}) \;  \RE G_{rr}(Q)^{\mu\nu} \hat{K}_{\mu} \hat{K}_{\nu}.
\end{equation}
This collision kernel also describes the gluon rungs. Defining 
\beq 
\widetilde{f}^{\mu}(\mathbf{p}) = -4p^z (p^z + k) \int \frac{dp^0}{2\pi} \mathcal{F} \mathcal{D}^{\mu}.
\eeq
one gets that
\beq \label{eq:integral_eq}
\sigma^{\mu} = i \delta E \widetilde{f}^{\mu}(\mathbf{p}) + \int \frac{d^2 q_{\perp}}{(2\pi)^2}\; \mathcal{C}(\mathbf{q}_{\perp}) \left[ \widetilde{f}^{\mu}(\mathbf{p}) - \widetilde{f}^{\mu}(\mathbf{p} + \mathbf{q}_{\perp})\right].
\eeq

The production rate of photons with momentum \(\mathbf{k}\) is 
\begin{widetext}
\begin{eqnarray}
\label{eq:photonrate2}
k \frac{dR}{d^3 k} = \frac{3Q^2 \alpha_{EM}}{4\pi^2} 
 \int \frac{d^3 p}{(2\pi)^3} F(P+K) \left[ 1- F(P)\right] \frac{p^{z\;2} + (p^z+k)^2}{2p^{z\;2} (p^z + k)^2} \;\mathbf{p}_{\perp} \cdot \RE \mathbf{f}(\mathbf{p};\mathbf{k})
\end{eqnarray}
\end{widetext}
\noindent as can be seen by using Eq. \eqref{eq:1122andaarr}, Eq. \eqref{eq:photonrate2} and evaluating the trace of the quark loop. Here \(\mathbf{f}\) is the transverse part of \(\widetilde{f}\) without the Pauli matrix.
The new factors in \(p^z\) and \(k\) come from summing over the physical polarization of the photon  \cite{Arnold2001}. Furthermore
\(Q\) is defined by 
\beq 
Q^2 e^2 = \sum_{\mathrm{flavour}} q^2
\eeq
where we sum over the different flavours of light quarks. 
As explained above,  \(F\) is the momentum distribution of quarks including Fermi suppression for outgoing quarks, 
\beq \label{eq:F(P)}
F(P) = f_q(\mathbf{p}) \theta(p^0) + (1-f_{\bar{q}}(-\mathbf{p})) \theta(-p^0)
\eeq
Finally \(p^0 = (-k^0 + E_{\mathbf{p}}\; \mathrm{sgn}(p^z) + E_{\mathbf{p} + \mathbf{k}} \; \mathrm{sgn}(p^z + k))/2\).

In this expression \(\mathbf{f}\) satisfies a Boltzmann-like integral equation
\beq 
\mathbf{p_{\perp}} = i \delta E\; \mathbf{f}(\mathbf{p_{\perp}}) + \int \frac{d^2 q_{\perp}}{(2\pi)^2}\; \mathcal{C}(\mathbf{q}_{\perp}) \left[\mathbf{f}(\mathbf{p_{\perp}}) - \mathbf{f}(\mathbf{p_{\perp}} + \mathbf{q}_{\perp})\right].
\label{eq:B-like}
\eeq
Here \(\mathbf{f}(\mathbf{p_{\perp}}; \;p^z, \mathbf{k})\)
 is an analog of the density of hard quarks with transverse momentum \(\mathbf{p}_{\perp}\) that emit a photon with momentum \(\mathbf{k}\). The term  \(i \delta E\; \mathbf{f}\) works like a time derivative in momentum space. The integral describes the change in transverse momentum of the quarks through the exchange of soft gluons with the medium. The gain term comes from the gluon rungs while the loss term comes from the quark decay width.  Eqs. \eqref{eq:photonrate} and \eqref{eq:B-like} agree with the abelian limit of the results of \cite{AMY_eff_kin} where a kinetic theory of quarks and gluons was used to study the LPM effect in a perhaps more heuristic way.

In an isotropic plasma, \(f(\mathbf{p}) = f(p)\),  one gets a simple expression for the collision kernel \(\mathcal{C}(\mathbf{q}_{\perp})\) by using a sum rule \cite{AMY_eff_kin,Aurenche2002}. This special case is relevant for the bulk viscous correction to photon production. Specifically,
\beq 
\mathcal{C}(\mathbf{q}_{\perp}) = g^2 C_F \; \Omega  \bigg[\frac{1}{\mathbf{q}_{\perp}^2} - \frac{1}{\mathbf{q}_{\perp}^2 + m^2_D}   \bigg].
\eeq
Here
\begin{equation} \label{eq:Omega}
\Omega = \frac{\int_0^{\infty} dp \,p^2\,   \left[ 2 N_f f_q (1-f_q)+ 2 N_c f_g (1+f_g)\right]}{-\int_0^{\infty} dp \,p^2\,    \frac{d}{d p}\left[2 N_f f_q + 2 N_c f_g \right]}
\end{equation}
characterizes the occupation density of soft gluons. Furthermore,
\beq
m_D^2 = \frac{g^2}{\pi^2}\int_0^{\infty} dp \; p \left[2 N_f f_q(p) + 2N_c f_g(p) \right]
\eeq
is a non-equilibrium Debye mass. In the isotropic case the LPM effect only depends on \(\Omega\), \(m_D^2\) and \(m_{\infty}^2\), the non-equilibrium mass of hard quarks, along with the momentum distribution \(F\) from Eq. \eqref{eq:F(P)}.

In an anisotropic plasma the collision kernel in Eq. \eqref{eq:coll_kern} and the quark decay width in Eq. \eqref{eq:decay_and_kernel} are divergent, because of the gauge field instabilities discussed previously. For the moment, the simplest solution is to impose that the anisotropy is small enough for the divergences to be subleading in the coupling. At leading order only \(g^2 T \lesssim Q \lesssim gT\) contributes to the kernel so one demands that the divergence takes place at the ultrasoft scale of \(Q \lesssim g^2 T\).
For the momentum distribution in Eq. \eqref{eq:RS_mom_distr} this leads to \(| \xi | \lesssim g^2\). We illustrate this with the  $\alpha$ collective mode of Ref. \cite{Romatschke2003}. In that case, the divergence in the quark decay width comes from terms like 
\beq
\label{eq:divergence}
\int^{gT}_{g^2T} \! d^4 Q\;G_{\ret} G_{\adv} \sim \int^{gT}_{g^2T} \frac{d^4 Q}{ \left(\mathbf{q}^2 + m_\alpha^2 \right)^2 + \left(\frac{\pi q^0}{4 q} m_D^2 \right)^2}
\eeq
where  \(G_{\adv} = G_{\ret}^*\) and we have taken the limit $q^0 \to 0$ where possible. At leading order in \(\xi\), the static limit of the \(\alpha\) self-energy component is \cite{Romatschke2003}
\beq
m_\alpha^2 = -\frac{\xi}{6} \left( 1 + \cos 2 \theta_n\right) m_D^2.
\eeq
It depends on the angle between the gluon momentum and the direction of the anisotropy, \(\theta_n\). For \(\xi>0\), \(m_\alpha^2\) is negative which leads to divergences. Clearly, the divergence is at \(q \lesssim g^2 T\) if \(\xi \lesssim g^2\). 
%Specifically, the pole makes the collision kernel defined in Eq. \eqref{eq:coll_kern} divergent. Throughout our analysis we will assume that the anisotropy is small enough so that this divergence does not appear at leading order. The collision kernel integrates momenta with \(g^2 T \lesssim Q \lesssim gT\).}

\vspace{0.5cm}
\section{Conclusion} \label{Conclusion}
The quark-gluon plasma created in heavy-ion collisions deviates from local thermal equilibrium. 
To understand how the plasma radiates photons one needs to include the effects of these deviations. This means analyzing photon production in a non-equilibrium plasma. 
 Understanding the effect of viscosity on photons could in turn be used to extract the transport coefficients of QGP from photonic observables.

In this work we have studied photon production through bremsstrahlung and pair annihilation in a non-equilibrium QGP. Using field theory, we derived  integral equations describing these channels and the LPM effect: Eqs. (\ref{eq:B-like}) and (\ref{eq:coll_kern}). Along the way, we showed that the resummed \(rr\) propagator of gluons in an anisotropic plasma has an imaginary part that does not contribute at leading order but could be important at higher order. We also derived a simple expression for the \(rr\) propagor of hard, on-shell quarks. Finally, we presented a way of summing up ladder diagrams without using the KMS condition which is only true in thermal equilibrium.  Our derivation of the integral equation is valid for low anisotropy. In the special case of an isotropic plasma the integral equation only depends on three non-equilibrium constants.

To solve the integral equation one needs to assume some momentum distribution \(f(\mathbf{p})\). Work is ongoing on evaluating the bulk and shear viscous corrections to photon production using  \(f(\mathbf{p})\) derived from kinetic theory. 
%That boils down to evaluating \(G_{rr}\) for soft gluons. Unfortunately, such a calculation is plagued by gauge field instabilities in anisotropic plasmas, the fact that soft gluon modes grow exponentially \cite{Mrowczynski2016}. We will address this problem elsewhere.
 In the future, these ideas could be used to analyze jets in a non-equilibrium plasma and to extend this work to higher anisotropy. This might allow for the  extraction of transport coefficients of QGP from jet observables.

\begin{acknowledgements}

We thank Stanis\l{}aw Mr\'{o}wczy\'{n}ski for useful discussions. This work was supported in part by the Natural Sciences and Engineering Research Council of Canada. S. H. acknowledges a scholarship from the Department of Physics of McGill University, and C. G. acknowledges support from the Canada Council for the Arts through its Killam Research Fellowship program. 

\end{acknowledgements}

\bibliography{references}

\end{document}